\documentclass[Journal,twoside,web]{ieeecolor}
\usepackage{generic}
\usepackage{cite}
\usepackage{amsmath,amssymb,amsfonts}
\usepackage{graphicx}
\usepackage{algorithm,algorithmic}
\usepackage{hyperref}
\hypersetup{hidelinks}
\usepackage{textcomp}
\usepackage{setspace}
\usepackage{subcaption}
\usepackage{tikz}
\usepackage{xcolor}
\usepackage{needspace}
\definecolor{mypink}{RGB}{255,105,180}
\usepackage{overpic}       
\newtheorem{theorem}{\bf Theorem}
\newtheorem{assumption}{Assumption}
\newtheorem{remark}{Remark}
\newtheorem{lemma}{Lemma}
\newtheorem{example}{Example}
\newtheorem{definition}{Definition}
\newtheorem{proposition}{Proposition}
\newtheorem{corollary}{Corollary}
\def\BibTeX{{\rm B\kern-.05em{\sc i\kern-.025em b}\kern-.08em
T\kern-.1667em\lower.7ex\hbox{E}\kern-.125emX}}
\begin{document}
\bstctlcite{IEEEexample:BSTcontrol}
\title{Input-to-State Safety and Small-Gain Theorems for Large-Scale Nonlinear Systems}

\author{Jianqiang Wang, Tengfei Liu, and Zhong-Ping Jiang
\thanks{This work was supported in part by NSFC grants 62325303, 62333004, and 62521001, and in part by NSF grant ECCS-2210320. (Corresponding author: T. Liu)}
\thanks{Jianqiang Wang and Tengfei Liu are with the State Key Laboratory of Synthetical Automation for Process Industries, Northeastern University,
Shenyang 110819, China (e-mail: 18524460249@163.com; tfliu@mail.neu.edu.cn). }
\thanks{Zhong-Ping Jiang is with the Department of Electrical and Computer Engineering, New York University, Brooklyn, NY 11201 USA (e-mail: zjiang@nyu.edu).}}

\maketitle
\begin{abstract}
This paper studies the safety verification problem for a general class of large-scale systems characterized by multiple safety indicator functions. Each safety indicator function satisfies a trajectory-based input-to-state safety (ISSf) estimate that provides a lower bound on its value along system trajectories in terms of the initial state, external inputs, and the other safety indicator functions, whose influence is quantified by continuous and nondecreasing gain functions. Within this framework, we establish a safety-oriented small-gain theorem and its corresponding corollaries that enable compositional and scalable safety verification of the interconnected system, provided that the interconnection gain functions satisfy a small-gain condition on the negative real axis. This condition concerns only the neighborhood of the safety boundary, thereby facilitating the application of our results to safety verification of multi-agent systems under distributed control. Our proof combines small-gain techniques with the lower-bound structure of ISSf estimates and develops an inductive comparison argument that establishes the ISSf property for the overall system throughout the maximal interval of existence. As an application, we establish robust safety for the natural distributed “half-spring” controller, theoretically supporting the computationally efficient and inspiring approach to safety-critical control of multi-agent systems in the presence of obstacles and external disturbances.
\end{abstract}


\section{Introduction}
\label{sec:introduction}

Considerable effort has been devoted to developing control-theoretic tools for constraint satisfaction in dynamical systems, with safety-critical control being a prominent application. Representative approaches include velocity-obstacle methods \cite{VESENTINI_2024_RAS}, reachability analysis \cite{Chen_2018_ARCRAS}, potential-function methods \cite{Duhe2021APF}, and control barrier functions \cite{Ames2019ECC}. These methods have been applied to a broad range of constraint-satisfaction problems, including systems with high relative degree \cite{Rufli_2013_TRO,Nguyen2016ACC_a,Xiao2022TAC_a,Breeden2023AUTO}, hybrid systems \cite{LYGEROS_1999_auto,Nguyen2016ACC}, and nonlinear systems subject to uncertainty \cite{VERGINIS_2021_AUTO,Jankovic2018AUTO,Cosner2021IROS,Alan2023TCST}. Their practical efficacy has also been demonstrated through experimental implementations in a variety of engineering systems, including robotic manipulators \cite{Singletary2019IROS}, humanoid robots \cite{Escande2014IJRR}, human–robot interaction systems \cite{Zacharaki2020SafetyScience}, and multi-agent systems \cite{Wang2017TRO}.

As dynamical systems increase in scale, there is a growing need for safety-verification tools whose complexity does not deteriorate with the number of subsystems. From a systems-engineering perspective, one would expect the safety of a large-scale system to be verifiable from the safety properties of its constituent subsystems and the structure of their interconnections. Such a constructive approach would be meaningful for many safety-verification problems in safety-critical control, including those arising in systems with high relative degree \cite{Wu_2026_TAC} and multi-agent systems \cite{Wu_2023_TAC}. However, subsystem-level safety does not, in general, guarantee safety of the interconnected system. A closely related paradigm has been established for the stability analysis and control of large-scale nonlinear systems. In particular, the nonlinear small-gain theorem \cite{dashkovskiy_2007_MCSS,Jiang_2008_WCICA} provides a systematic means of establishing stability for interconnected systems of input-to-state stable (ISS) or input-to-output stable (IOS) subsystems by verifying a small-gain condition on the interconnection gain functions.

Small-gain techniques have subsequently been extended to safety-critical systems \cite{LYU_2022_auto,Lyu_2024_IEEETAC,WANG_2025_IFAC}, providing tools for safety verification of interconnected ISSf systems characterized by gradient-based barrier functions. In particular, the results in \cite{LYU_2022_auto,Lyu_2024_IEEETAC} show that the interconnected system possesses the desired safety properties if the interconnection gain functions satisfy a small-gain condition on both the positive and negative real axes. The work in \cite{WANG_2025_IFAC} constructs an ISSf-barrier function for the large-scale system under a small-gain condition imposed only on the negative real axis, with additional regularity assumptions on the gain functions. However, many systems are naturally described in terms of their trajectories, while satisfying the small-gain condition on both the positive and negative real axes may be challenging in practical control applications.

Specifically motivated by safety-critical control of multi-agent systems, this paper considers a general class of large-scale systems which admit multiple safety indicator functions. Each safety indicator function satisfies a trajectory-based ISSf estimate that provides a lower bound on its value along system trajectories in terms of the initial state, external inputs, and the other safety indicator functions. The influence of other safety indicator functions on this lower bound is quantified by continuous and nondecreasing gain functions. This framework is sufficiently general to accommodate multi-agent systems subject to simultaneous collision-avoidance constraints between agents and between agents and obstacles. It also covers the case where the large-scale system can be decomposed into subsystems, each associated with a safety indicator function satisfying the gradient-based conditions for an ISSf-barrier function.

The main contribution of this paper is the development of a scalable, easily verifiable small-gain theorem that enables safety verification of large-scale systems by directly characterizing the magnitude-like dynamic interconnections among safety indicator functions. Our main result establishes that a large-scale system possesses the desired safety properties with respect to all safety indicator functions provided that the composition of the gain functions along every simple cycle in the interconnection graph is greater than the identity function on the negative real axis. The region on which this small-gain condition is imposed corresponds to the critical region for safety verification, namely, the neighborhood of the safety boundary. Therefore, our result applies to a broader class of large-scale nonlinear control systems, such as multi-agent systems in which safety-critical controllers intervene only when agents are close to obstacles or to one another.

We build on key ideas and techniques from existing small-gain results \cite{Jiang_1994_MCSS,dashkovskiy_2007_MCSS,karafyllis_2011_vector} and formulate the analysis in terms of lower bounds on safety indicator functions to prove the main result. Specifically, we exploit the lower-bound structure of the ISSf estimates and employ an inductive comparison argument to construct an increasing sequence of time instants and derive successive lower bounds on the safety indicator functions. This construction enables us to establish the ISSf property of the overall system throughout its maximal interval of existence. If the overall system is forward complete, then the ISSf estimate holds over the entire timeline.

As a nontrivial application of the main theorem, we establish robust safety for a class of natural distributed “half-spring” controllers for multi-agent systems \cite{VIRAGH_2014_BB}. These distributed controllers augment command velocities with repulsive terms that behave like springs under compression: each term is activated only when the distance to an obstacle or another agent falls below a prescribed threshold, and its magnitude increases monotonically with the resulting distance shortfall. Although the effectiveness of these controllers has been demonstrated experimentally \cite{VIRAGH_2014_BB,VASARHELYI_2014_IROS,VASARHELYI_2018_SciRob}, to the best of our knowledge, their robust safety properties have not been theoretically established in the existing literature.

The rest of the paper is organized as follows. Section \ref{Preliminaries} introduces notations and the notion of ISSf. Section \ref{section_Motivating_Example} presents a motivational example of multi-agent safety-critical control and characterizes the interactions among the safety indicator functions under a class of natural distributed “half-spring” controllers. Section \ref{section_mainresult} formulates the general system framework and presents the main small-gain theorem and its corollaries. Section \ref{section_proof_mainresult} provides the proof of the main theorem. Section \ref{section_apply} applies the theorem to establish robust safety of the natural distributed “half-spring” controllers in the presence of external disturbances and presents numerical simulation results. Section \ref{section_conclude} concludes the paper. A technical lemma and supporting proofs are provided in the appendix.

\Needspace{5\baselineskip}

\section{Preliminaries}
\label{Preliminaries}

\subsection{Notations}

In this paper, we use $|x|$ to represent the Euclidean norm of $x\in\mathbb{R}^n$. We denote $\mathbb{R}_{+}=[0,+\infty)$, $\mathbb{R}_{-}=(-\infty,0]$, and $\mathbb{Z}_+=\{0,1,2,\ldots\}$. For $\mathbf{a}=[a_1,\ldots,a_n]^T\in\mathbb{R}^n$, denote $\lfloor \mathbf{a}\rfloor=\min\{a_1,\ldots,a_n\}$. For a measurable and locally essentially bounded function $f:(b,c)\to\mathbb{R}^n$, we denote the essential supremum of $|f(s)|$ over the interval $J \subseteq (b,c)$ by $\| f\|_{J}$ (i.e., $\|f\|_{J}=\operatorname{ess\,sup}_{s\in J}|f(s)|$). For a measurable and locally essentially bounded function $f:(b,c)\to\mathbb{R}^n$, we denote the essential infimum of $\lfloor f(s)\rfloor$ over the interval $J$ by $\lfloor\!\lfloor f\rfloor\!\rfloor_J$ (i.e., $\lfloor\!\lfloor f\rfloor\!\rfloor_J=\operatorname{ess\,inf}_{s\in J}\lfloor f(s)\rfloor$). Moreover, we denote $\lfloor\!\lfloor f\rfloor\!\rfloor_{\emptyset}=0$.

For two real vectors $\mathbf{u}=[u_1,\ldots,u_n]^T\in\mathbb{R}^n$ and 
$\mathbf{v}=[v_1,\ldots,v_n]^T\in\mathbb{R}^n$, we say $\mathbf{u}\le \mathbf{v}$ if $u_i\le v_i$ for 
$i=1,\ldots,n$, and $\mathbf{u}<\mathbf{v}$ if $\mathbf{u}\le \mathbf{v}$ and $\mathbf{u}\neq \mathbf{v}$. Define $\operatorname{MIN}\{\mathbf{u},\mathbf{v}\}=(\min\{u_1,v_1\},\ldots,\min\{u_n,v_n\})$. For any positive integer $n$, we denote by $\mathbf{0}_n$ and $\mathbf{1}_n$ the $n$-dimensional vectors of zeros and ones, respectively, and $\text{Id}_n(\mathbf{s})=\mathbf{s}$ for $\mathbf{s}\in\mathbb{R}^n$. For two operators $T_1:\mathbb{R}^n\to\mathbb{R}^m$ and $T_2:\mathbb{R}^m\to\mathbb{R}^p$, their composition $T_2 \circ T_1$ is defined by $(T_2 \circ T_1)(\mathbf{s})=T_2(T_1(\mathbf{s}))$ for $\mathbf{s}\in\mathbb{R}^n$. For a function $\Gamma:\mathbb{R}^n\to\mathbb{R}^n$, 
$[\Gamma]^i$ denotes the composition $\underbrace{\Gamma \circ \cdots \circ \Gamma}_{i}$, and $[\Gamma]^0=\text{Id}_n$. 

Definitions of class $\mathcal{K}$, class $\mathcal{K}_{\infty}$, and class $\mathcal{KL}$ functions  can be found in standard texts \cite{khalil_2002_nonlinear}. A function $\gamma:\mathbb{R}\rightarrow\mathbb{R}$ is said to be of class $\mathcal{K}^-$, denoted by $\gamma\in\mathcal{K}^-$, if there exists a $\gamma^+\in\mathcal{K}$ such that $\gamma(s)=-\gamma^+(-s)$ for all $s<0$ and $\gamma(s)=0$ for all $s\ge 0$. A function $\gamma\in\mathcal{K}^-$ is said to be of class $\mathcal{K}^-_\infty$, if $\lim_{s\to-\infty}\gamma(s)=-\infty$. A function $\gamma:\mathbb{R}\rightarrow\mathbb{R}$ is said to be of class $\mathcal{K}^e$, denoted by $\gamma\in\mathcal{K}^e$, if it is continuous, strictly increasing, and $\gamma(0)=0$. A function $\alpha : \mathbb{R} \to \mathbb{R}$ is said to be of class $\mathcal{M}^e$, if it is continuous and satisfies $\alpha(0)=0$ and $s\alpha(s)>0$ for all $s\neq 0$. A continuous function $\beta: \mathbb{R}\times \mathbb{R}_+\to \mathbb{R}$ is said to be of class $\mathcal{K}^-\mathcal{L}$ if, for each fixed $t\ge 0$, $\beta(\cdot,t)$ is of class $\mathcal{K}^-$, and for each fixed $s<0$, $\beta(s,\cdot)$ is increasing and tends to zero as $t\to +\infty$.

For a real-valued function $f:(a,b)\to\mathbb{R}$, the upper right Dini derivative at $t\in(a,b)$ is defined as
\begin{align}
D^{+}f(t) 
= \limsup_{h\to 0^{+}} \frac{f(t+h)-f(t)}{h}.
\end{align}
For a vector-valued function
$f=[f_1,\ldots,f_n]^T:(a,b)\to\mathbb{R}^n$,
the upper right Dini derivative at $t\in(a,b)$ is defined
componentwise as
\begin{align}
D^{+}f(t)
&=[D^{+}f_1(t),\ldots,D^{+}f_n(t)]^T.
\end{align}
Moreover, for a function $g:\mathcal{D}\to\mathbb{R}$, where $\mathcal{D}\subseteq\mathbb{R}^n$ is an open set, the upper right Dini derivative at $x\in\mathcal{D}$ in the direction $v\in\mathbb{R}^n$ is defined as
\begin{align}
D^{+}g(x;v)=\limsup_{t\to 0^{+}}\frac{g(x+tv)-g(x)}{t}.
\end{align}

\subsection{Input-to-State Safety (ISSf)}

Consider a system
\begin{align}
\dot{x}=f(x,w)\label{preliminary_system_dynamics}
\end{align}
where $x\in\mathbb{R}^n$ is the state, $w\in\mathbb{R}^m$ is the external input, and $f:\mathbb{R}^n\times\mathbb{R}^m\to\mathbb{R}^n$ denotes the dynamics of the system, which is locally Lipschitz continuous on $\mathcal{D}\times\mathbb{R}^m$ with $\mathcal{D}\subseteq\mathbb{R}^n$ being an open set. To simplify the notation, $w : \mathbb{R}_{+} \to \mathbb{R}^{m}$ is also used to denote a function of time. In that case, it is assumed to be measurable and locally essentially bounded.

For any $x(0)\in\mathcal{D}$ and any measurable and locally essentially bounded $w$, there exists a unique maximal Carath\'eodory solution $x(t)$ to system \eqref{preliminary_system_dynamics}, defined on $[0,t_m)$ \cite[Theorems I.5.1 and I.5.2]{hale1980ode}, with $t_m\in(0,\infty]$ being the maximal existence time of the solution. Moreover, if $t_m<\infty$, then $x(t)$ approaches the boundary of $\mathcal{D}$ as $t\to t_m$.

\begin{definition}\label{definition_ISSf}
System \eqref{preliminary_system_dynamics} is said to be ISSf with respect to a continuous function $h:\mathbb{R}^n\rightarrow\mathbb{R}$ if there exist functions $\beta\in\mathcal{K}^-\mathcal{L}$, $\gamma\in\mathcal{K}^-$, continuous functions $\sigma:\mathcal{D}\to\mathbb{R}$ and $\psi:\mathbb{R}^m\to\mathbb{R}$, and a constant $c\le 0$ such that, for any initial state $x(0)\in\mathcal{D}$ and any measurable and locally essentially bounded external input $w:\mathbb{R}_{+}\to\mathbb{R}^m$,
\begin{align}
h(x(t))\ge \min\{\beta(\sigma(x(0)),t),\gamma(\lfloor\!\lfloor \psi(w(\cdot))\rfloor\!\rfloor_{[0,t)}),c\}
\label{tra_ISSf}
\end{align}
holds for all $t\in[0,t_m)$. The function $h$ is called a safety indicator function.
\end{definition}

\begin{remark}
Definition \ref{definition_ISSf} extends the notions of ISSf introduced in \cite{Romdlony_2016_CDC,Kolathaya_2019_IEEECSL}. In \cite{Romdlony_2016_CDC}, system \eqref{preliminary_system_dynamics} is said to be ISSf with respect to a set of unsafe states $\mathcal{Z}\subset \mathbb{R}^n$ if there exist $\alpha,\phi:\mathbb{R}_{+}\times\mathbb{R}_{+} \to\mathbb{R}_{+}$ and $\varphi\in\mathcal{K}$ such that
\begin{align}
\varphi(h(x(t)))\ge \alpha(h(x(0)),t)-\phi(\|w\|_{[0,t)},t)
\end{align}
for all $t\ge 0$, where $h(x)=\inf_{a\in\mathcal{Z}}|x-a|$ for $x\in\mathbb{R}^n$ represents the distance from $x$ to the set $\mathcal{Z}$, and $\alpha(\cdot,t),\phi(\cdot,t)\in\mathcal{K}$ for fixed $t\ge 0$. In \cite{Kolathaya_2019_IEEECSL}, system \eqref{preliminary_system_dynamics} is said to be ISSf with respect to a set $\mathcal{C}:=\{x\in\mathbb{R}^n:h(x)\ge 0\}$, if there exists $\gamma_w\in\mathcal{K}$ such that the set $\mathcal{C}_w=\{x\in\mathbb{R}^n:h(x)+\gamma_w(\|w\|_{[0,\infty)})\ge 0\}$ is forward invariant, where $h:\mathbb{R}^n\to\mathbb{R}$ is a continuously differentiable function. In accordance with \cite{Romdlony_2016_CDC,Kolathaya_2019_IEEECSL}, Definition \ref{definition_ISSf} characterizes the effects of both the initial state and the external input on safety. 
Slightly different from the definitions in \cite{Romdlony_2016_CDC,Kolathaya_2019_IEEECSL}, our definition uses $\lfloor\!\lfloor\psi(w)\rfloor\!\rfloor_{[0,t)}$ instead of $\|w\|_{[0,t)}$ or $\|w\|_{[0,\infty)}$, to accommodate cases where the safety indicator function is affected by only a subset of the external input components. Moreover, based on our definition, the set $\mathcal{C}_w=\{x\in\mathbb{R}^n: h(x)+\gamma_w(\|w\|_{[0,\infty)})\ge 0\}$ is not required to be forward invariant, which allows the trajectory of the safety indicator function $h(x(\cdot))$ to exhibit overshoots. As shown by the example of multi-agent systems in Section \ref{section_Motivating_Example}, the refined definition effectively captures the natural ``half-spring'' interdependence among the controlled agents' safety properties, which is vital for the safety analysis of the practically meaningful multi-agent system.
\end{remark}

\begin{remark}
Note that $h(x(0))$ may not fully capture the influence of the initial state $x(0)$. This is the reason why we use $\sigma(x(0))$ instead of $h(x(0))$ to define ISSf in \eqref{tra_ISSf}. The effectiveness of such treatment is illustrated in Example \ref{example_ISSf}. In addition, we employ the constant $c$ as a practical term to characterize the influence of perturbations on safety. An example can be found in Section \ref{section_Motivating_Example}.
\end{remark}

Example \ref{example_ISSf} illustrates the ISSf property of a simple second-order linear system in the sense of Definition \ref{definition_ISSf}.

\begin{example}\label{example_ISSf}
Consider the system
\begin{align}
\dot{x}_1=x_2,\qquad
\dot{x}_2=-x_1-2x_2+w,
\label{example_equation}
\end{align}
where $x=[x_1,x_2]^T\in\mathbb{R}^2$ is the state and
$w:\mathbb{R}_+\to\mathbb{R}$ is the measurable and locally essentially
bounded disturbance input. Consider the safety indicator function
\begin{align}
h(x)=x_1.
\label{example_h}
\end{align}
Applying basic tools of linear systems theory \cite{Hespanha2018}, one can verify the existence of constants $M\ge1$ and
$\lambda>0$ such that
\begin{align}
&h(x(t))\nonumber\\
&\ge 2\min\left\{-Me^{-\lambda t}|x(0)|,\lfloor\!\lfloor w\rfloor\!\rfloor_{[0,t)}\int_0^t [1,0]e^{A(t-\tau)}B\,d\tau\right\}\nonumber\\
&\ge 2\min\left\{-Me^{-\lambda t}|x(0)|,\lfloor\!\lfloor w\rfloor\!\rfloor_{[0,t)}\right\}
\end{align}
for all $t\ge 0$, with
\begin{align}
A=\begin{bmatrix}
0 & 1\\
-1 & -2
\end{bmatrix},
\quad
B=\begin{bmatrix}
0\\
1
\end{bmatrix}.
\end{align}
Thus, system \eqref{example_equation} is ISSf with respect to $h$ defined in \eqref{example_h}, with $w$ as the input. Specifically, property \eqref{tra_ISSf} holds with $\beta(s,t)=2Me^{-\lambda t}\min\{s,0\}$, $\gamma(s)=2\min\{s,0\}$, $\sigma(x)=-|x|$, $\psi(s)=s$ and $c=0$ for $x\in\mathbb{R}^2$, $s\in\mathbb{R}$ and $t\ge 0$.
\end{example}

The following theorem shows that system \eqref{preliminary_system_dynamics} is ISSf in the sense of Definition \ref{definition_ISSf} if the safety indicator function $h$ satisfies the gradient condition for ISSf barrier functions \cite[Definition 4]{Kolathaya_2019_IEEECSL} with the influence of external inputs $w$ characterized by $\tilde{\psi}(w)$.

\begin{theorem}\label{ISSf-BFtheorem}
If $h$ is locally Lipschitz continuous on $\mathcal{D}$, and there exist $\alpha\in\mathcal{M}^e$, $\chi\in\mathcal{K}^-$, and a continuous function $\tilde{\psi}:\mathbb{R}^m\to\mathbb{R}$ such that for any $x\in\mathcal{D}$ and $w\in\mathbb{R}^m$,
\begin{align}
h(x)\le \chi(\tilde{\psi}(w))\Rightarrow D^+h(x;f(x,w)) \ge -\alpha(h(x)),\label{definition_ISSfBF}
\end{align}
then system \eqref{preliminary_system_dynamics} is ISSf with respect to $h$, and there exists $\beta\in\mathcal{K}^-\mathcal{L}$ such that property \eqref{tra_ISSf} holds with $\gamma=\chi$, $\sigma=h$, $\psi=\tilde{\psi}$, and $c=0$. Moreover, $\beta$ satisfies $\beta(s,0)=s$ for all $s\le 0$.
\end{theorem}

\begin{proof}
For any $0\le T_1<T_2<t_m$, the solution $x$ is
absolutely continuous on $[T_1,T_2]$, and the set
$K=\{x(t):t\in[T_1,T_2]\}$ is a compact subset of
$\mathcal{D}$.
The local Lipschitz continuity of $h$ on $\mathcal{D}$
implies that its restriction to $K$ is Lipschitz continuous. Following the proof of \cite[Proposition 1.49]{Mironchenko_2023_book},
we conclude that $h(x(\cdot))$ is absolutely continuous on $[T_1,T_2]$. Under condition \eqref{definition_ISSfBF}, for any $x(0)\in\mathcal{D}$ and any measurable and locally essentially bounded $w:\mathbb{R}_+\to\mathbb{R}^m$,
\begin{align}
h(x(t))\le \chi(\tilde{\psi}(w(t)))\Rightarrow D^+h(x(t))\ge -\alpha(h(x(t)))\label{lemma1_condition}
\end{align}
holds for almost all $t\in [0,t_m)$.

Define
$\bar{h}(x)=-h(x)$, $\bar{\alpha}(r)=-\alpha(-r)$, $\bar{\psi}(\xi)=-\tilde{\psi}(\xi)$ and $\bar{\chi}(r)=-\chi(-r)$ for $x\in\mathbb{R}^n$, $\xi\in\mathbb{R}^m$, and $r\in\mathbb{R}$. We can rewrite \eqref{lemma1_condition} as
\begin{align}
\bar{h}(x(t))\ge \bar{\chi}(\bar{\psi}(w(t)))\Rightarrow D^+ \bar{h}(x(t))\le -\bar{\alpha}(\bar{h}(x(t)))
\end{align}
for almost all $t\in [0,t_m)$. Then, Theorem \ref{ISSf-BFtheorem} can be proved following the same lines as in \cite[Theorem 2.12]{Mironchenko_2023_book}.
\end{proof}

\section{A Motivational Example of Safety-Critical Control of Multi-Agent Systems}\label{section_Motivating_Example}

In this section, we consider a popular safety-critical control scenario involving $N$ mobile agents and $N_o$ obstacles. Related problems have been studied in the robotics and control community \cite{Wang2017TRO,Long_2017_RAL,Mestres_2024_RAL,Pham_2024_RAL}. With each mobile agent modeled as a first-order integrator subject to external disturbances, we consider an intuitive class of distributed safety-critical controllers as proposed in \cite{VIRAGH_2014_BB}, which are designed as the sum of lower-saturated relative distances, and are called ``half-spring'' controllers. The effectiveness of this class of controllers has also been experimentally demonstrated with specific controller parameter settings in \cite{VASARHELYI_2014_IROS,VASARHELYI_2018_SciRob}. However, we cannot find a dedicated systematic analysis of the robust safety of the resulting closed-loop system. There is also a lack of an explicit condition on the controller parameters that ensure safety. Indeed, for specific scenarios, counterexamples can be found; see Section \ref{subsection_numerical_Example}. From the viewpoint of interconnected systems, one technical difficulty for theoretical analysis is caused by the interdependence and interaction between the inter-agent and agent-obstacle relative positions. We define safety indicator functions based on the relative positions to explicitly characterize the interdependence and interaction, which motivates the study of this paper.

For $i=1,\ldots,N$, each agent $i$ is modeled as
\begin{align}
\dot{p}_i=u_i+w_i,\label{apply_system_oneorder}
\end{align}
where $p_i\in\mathbb{R}^2$ is the position, $u_i\in\mathbb{R}^2$ is the control input, and $w_i\in\mathbb{R}^2$ represents the external disturbance acting on agent $i$. To simplify the notation, for each $i=1,\ldots,N$, $w_i : \mathbb{R}_{+} \to \mathbb{R}^2$ is also used to denote a function of time. In that case, it is assumed to be measurable and essentially bounded. 

\begin{figure}[h!]
\centering
\begin{overpic}[width=0.8\linewidth]{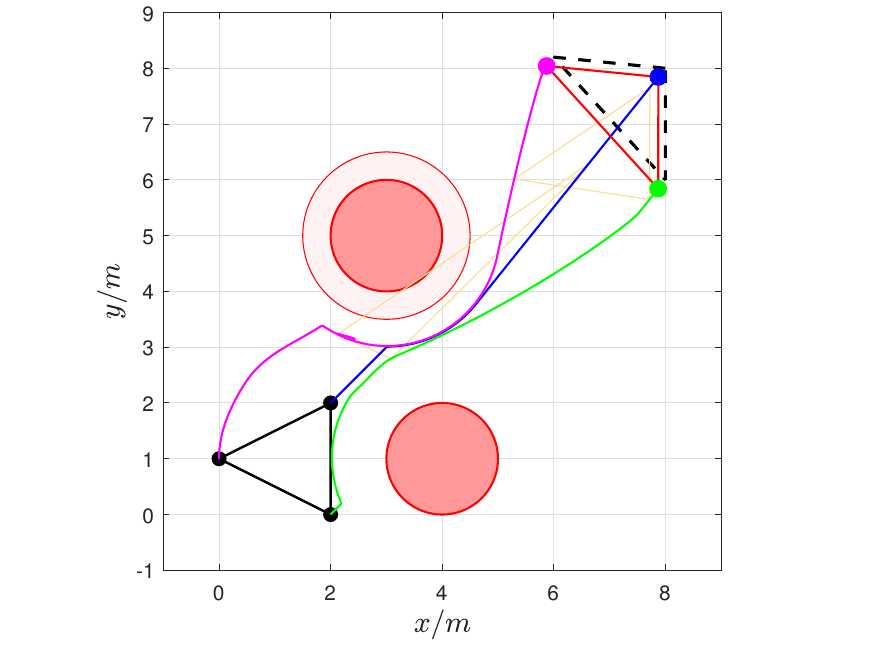}
\put(64,70){\scriptsize $\text{Destination}$}
\put(30,31){\scriptsize $\text{Agent 1}$}
\put(20,16){\scriptsize $\text{Agent 2}$}
\put(33,12){\scriptsize $\text{Agent 3}$}
\put(30,61){\scriptsize $\text{Obstacle 1}$}
\put(59,22){\scriptsize $\text{Obstacle 2}$}
\put(85,59){\scriptsize $\text{Current Points}$}
\put(-5,25){\scriptsize $\text{Starting Points}$}
\linethickness{0.4pt}
\put(42.5,48){
\begin{tikzpicture}
\draw[<->,black, thin] (0,0) -- (0,0.68);
\end{tikzpicture}
}
\put(45,52){\scriptsize $d_{mo}$}
\put(37.6,43.4){
\begin{tikzpicture}
\draw[<->,black, thin] (0,0) -- (0.35,0.35);
\end{tikzpicture}
}
\put(42.5,43.5){\scriptsize $d_{o}$}
\put(56.7,53){
\begin{tikzpicture}
\draw[->,mypink, thick] (0,0) -- (0.0005,0.0019);
\end{tikzpicture}
}
\put(62.2,52){
\begin{tikzpicture}
\draw[->,blue, thick] (0,0) -- (0.0005,0.0007);
\end{tikzpicture}
}
\put(65.5,46.2){
\begin{tikzpicture}
\draw[->,green, thick] (0,0) -- (0.0005,0.0004);
\end{tikzpicture}
}
\put(17,26){\vector(1,0){13}}   

\linethickness{0.4pt}
\put(84,60){\vector(-1,0){9}}   
\end{overpic}
\caption{Safety-critical control of multiple mobile agents: Three mobile agents and two obstacles.}
\label{fig1}
\end{figure}

For each $j=1,\ldots,N_o$, we use $o_j\in\mathbb{R}^2$ to represent the position of the obstacle $j$. In this scenario, each agent $i$ is required to avoid collisions with the obstacles and the other agents, i.e., $|p_i - o_j| \ge d_o$ for $j=1,\ldots,N_o$ and $|p_i-p_j|\ge d_a$ for $j\neq i$, while trying to track a command velocity $v_i^c:\mathbb{R}_{+}\to\mathbb{R}^2$. Here, the constant $d_o>0$ represents the desired minimum distance between the agents and the obstacles, and the constant $d_a> 0$ represents the desired minimum distance between any two agents. To simplify the notation, denote 
\begin{align}
p=[p_1^T,\ldots,p_N^T]^T,\quad w=[w_1^T,\ldots,w_N^T]^T.
\end{align}
Figure \ref{fig1} illustrates a specific scenario of three mobile agents and two obstacles.

To quantify the safety of the agents, we define the following safety indicator functions
\begin{align}
h_i(p)&=\min_{j=1,\ldots,N_o}\{|p_i-o_j|-d_{mo}\},\quad i=1,\ldots,N,\label{definition_h_motivalexample}\\
h_{ij}(p)&=|p_i-p_j|-d_{ma},\quad i,j=1,\ldots,N,~i\neq j\label{definition_h_ij_motivalexample}
\end{align}
for $p\in\mathbb{R}^{2N}$, where the constants $d_{mo}>d_o$ and $d_{ma}>d_a$ represent the safety margins between the obstacles and the mobile agents and between the mobile agents, respectively. 

For each $i=1,\ldots,N$, consider the following controller for agent $i$:
\begin{align}
u_i&=-k_s v^s_i(p)-k_s \lambda\sum_{\substack{j=1,\ldots,N \\ j\neq i}}v^s_{ij}(p)+v_i^c\label{apply_ui_definition_N}
\end{align}
for all $p\in\mathcal{D}$, with
\begin{align}
v^s_{i}(p)&=\begin{cases}
\frac{\partial h_i(p)}{\partial p_i}h_i(p), & h_i(p)\le 0,\\
0,& h_i(p)> 0,
\end{cases}\label{definition_v_s_i}\\
v^s_{ij}(p)&=\begin{cases}
\frac{\partial h_{ij}(p)}{\partial p_i}h_{ij}(p), & h_{ij}(p)\le 0,\\
0,& h_{ij}(p)> 0,
\end{cases}\label{definition_v_s_ij}
\end{align}
constants $k_s>0$, $\lambda\in(0,1)$, and
\begin{align}
\mathcal{D}&=\{p\in\mathbb{R}^{2N}:h_i(p)> -d_{mo},h_{ij}(p)> -d_{ma},\nonumber\\
&~~~\forall i,j=1,\ldots,N,j\neq i\}.\label{apply_D_definition}
\end{align}
  
The following proposition gives an intuitive expression of the controller \eqref{apply_ui_definition_N}.

\begin{proposition}\label{proposition_controller_equivalence}
If the positions of obstacles satisfy
\begin{align}
|o_i-o_j|>2d_o, \quad i,j=1,\ldots,N_o,\; i\neq j,\label{condition_obstacle}
\end{align}
and the parameter $d_{mo}$ is chosen to satisfy
\begin{align}
2d_o<2d_{mo}<|o_i-o_j|,\quad i,j=1,\ldots,N_o,\;i\neq j,\label{d_mo_condition_1}
\end{align}
then, for any $p\in\mathcal{D}$, the controller \eqref{apply_ui_definition_N} is equivalent to
\begin{align}
u_i&=-k_s\sum_{j=1}^{N_o}\frac{p_i-o_j}{|p_i-o_j|}\min\{|p_i-o_j|-d_{mo},0\}\nonumber\\
&~~~-k_s \lambda\sum_{\substack{j=1,\ldots,N \\ j\neq i}}\frac{p_i-p_{j}}{|p_i-p_j|}\min\{|p_i-p_j|-d_{ma},0\}+v_i^c.\label{apply_ui_definition_N_equivalent}
\end{align}
\end{proposition}

The proof of Proposition \ref{proposition_controller_equivalence} is given in Appendix \ref{proof_proposition_controller_equivalence}. 

The first two terms in the controller \eqref{apply_ui_definition_N_equivalent} enforce the agent-obstacle and inter-agent safety constraints, respectively, and have the same form as the half-spring controller in \cite{VIRAGH_2014_BB,VASARHELYI_2014_IROS,VASARHELYI_2018_SciRob}. The third term, $v_i^c$, serves as a feedforward term for the command velocity to guide agent $i$ toward its motion objective, and such command velocities are used for target tracking, formation control, and flocking in \cite{VIRAGH_2014_BB,VASARHELYI_2014_IROS,VASARHELYI_2018_SciRob}.

The following proposition shows that the safety indicator functions $h_i$ and $h_{ij}$ possibly interact with each other, and their dynamic behaviors satisfy ISSf estimates as given by \eqref{tra_ISSf}. 

\begin{proposition}\label{prop_robust_safety_controller}
Consider the multi-agent system \eqref{apply_system_oneorder} with the controller \eqref{apply_ui_definition_N}. If the positions of the obstacles satisfy \eqref{condition_obstacle}, and there exists a constant $\bar{v}_c\ge 0$ such that
\begin{align}
|v_i^c(t)|\le \bar{v}_c,\quad i=1,\ldots,N\label{v_c_condition}
\end{align}
for $t\ge 0$, then, provided that the controller parameters $d_{mo}$, $d_{ma}$, and $\lambda$ are chosen to satisfy \eqref{d_mo_condition_1} and
\begin{align}
d_{ma}>\frac{1+N\lambda}{2\lambda}d_a,\label{d_ma_condition}
\end{align}
for each $i,j=1,\ldots,N$, $j\neq i$, there exist $\beta_o,\beta_a\in\mathcal{K}^-\mathcal{L}$, $\hat{\gamma}_{ik},\chi_o,\chi_a\in\mathcal{K}^-$ for $k\neq i$, $c_o\le 0$, and $c_a\le 0$ such that for any $p(0)\in\mathcal{D}$, any measurable $v_i^c$ satisfying \eqref{v_c_condition}, and any measurable and essentially bounded $w:\mathbb{R}_{+}\to\mathbb{R}^{2N}$,
\begin{align}
h_i(p(t))&\ge \min\{\beta_o(h_i(p(0)),t),\nonumber\\
&~~~\min_{\substack{k=1,\ldots,N\\ k\neq i}}\hat{\gamma}_{ik}(\lfloor\!\lfloor h_{k}(p(\cdot))\rfloor\!\rfloor_{[0,t)}),\nonumber\\
&~~~\chi_o(-\|w_i\|_{[0,t)}),c_o\}\label{interconnection_N}\\
h_{ij}(p(t))&\ge \min\{\beta_a(h_{ij}(p(0)),t),\chi_a(-\|w_i\|_{[0,t)}),\nonumber\\
&~~~\chi_a(-\|w_j\|_{[0,t)}),c_a\}\label{h_ij_robust}
\end{align}
for all $t\in[0,t_m)$, where $[0,t_m)$ is the time interval over which $p(t)$ is right-maximally defined. Moreover, $\beta_o(s,0)=\beta_a(s,0)=s$ for all $s\le 0$.
\end{proposition} 

\begin{figure}[h!]
\centering
\includegraphics[width=0.9\linewidth]{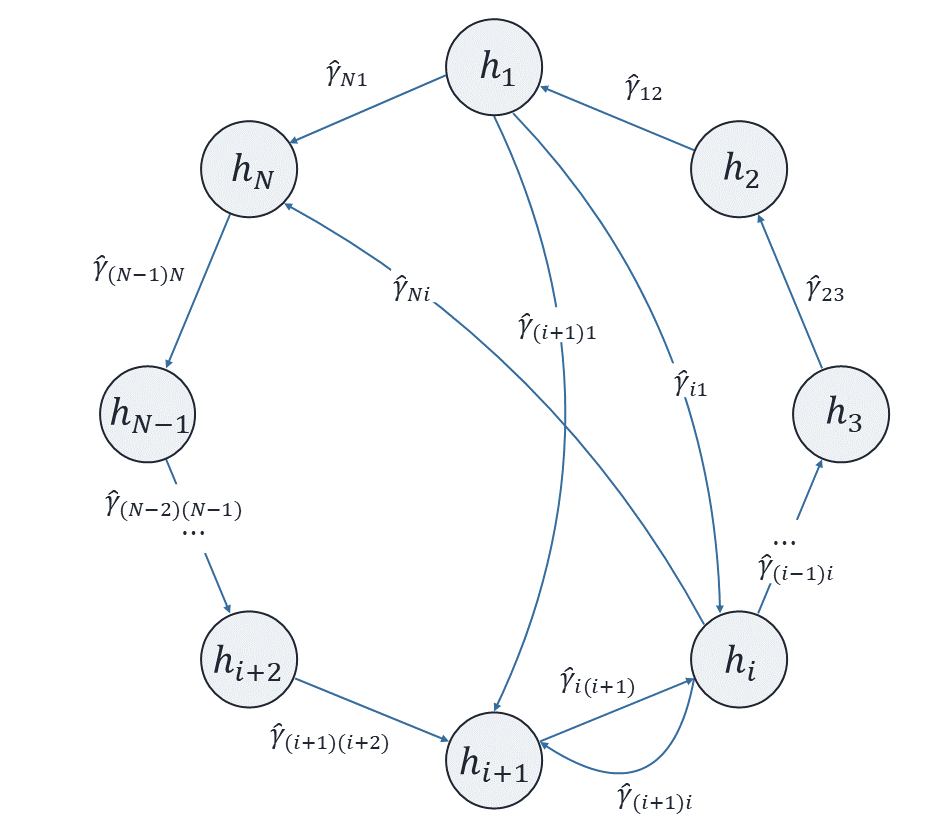}
\caption{Schematic illustration of some of the interconnections among the safety indicator functions.}
\label{fig_gaininterconnection}
\end{figure}

The proof of Proposition \ref{prop_robust_safety_controller} is given in Appendix \ref{proof_prop_robust_safety_controller}. It can be observed that $h_i(p(t))$ admits a lower bound that does not explicitly depend on $h_{ij}(p(t))$, because the contribution of $h_{ij}(p(t))$ to the derivative of $h_i(p(t))$ admits a lower bound depending only on the safety indicator functions $h_k(p(t))$, $k=1,\ldots,N$, describing the agent-obstacle safety. Likewise, the lower-bound estimate for $h_{ij}(p(t))$ does not explicitly depend on the safety indicator functions $h_i(p(t))$ and $h_j(p(t))$ describing the agent-obstacle safety or the other safety indicator functions $h_{ik}(p(t))$ and $h_{jk}(p(t))$ for $k\neq i,j$ describing the inter-agent safety, since their combined contribution to the derivative of $h_{ij}(p(t))$ is bounded below by a function of $|p_i(t)-p_j(t)|=h_{ij}(p(t))+d_{ma}$ alone. Some of the interconnections among the safety indicator functions are shown in Figure \ref{fig_gaininterconnection}. It is therefore natural to ask under what conditions the safety of all the interacting agents can be ensured simultaneously. This motivates the study in this paper.

\Needspace{5\baselineskip}

\section{System Description and Main Result}\label{section_mainresult}

\subsection{System Description}

In this paper, we consider nonlinear systems described by
\begin{align}
\dot{x}=f(x,w),\label{network_dynamics}
\end{align}
where $x \in \mathbb{R}^{n}$ is the state, 
$w \in \mathbb{R}^{m}$ represents external disturbance inputs, and $f:\mathbb{R}^n\times\mathbb{R}^m\to\mathbb{R}^n$ denotes the dynamics of the system, which is locally Lipschitz continuous on $\mathcal{D}\times\mathbb{R}^m$ with $\mathcal{D}\subseteq\mathbb{R}^n$ being an open set. To simplify the notation, $w$ is also used to denote a function of time $w : \mathbb{R}_{+} \to \mathbb{R}^m$. In that case, it is assumed to be measurable and essentially bounded. 

Consider a general case with $N$ safety indicator functions $h_i:\mathbb{R}^n\to\mathbb{R}$ for $i=1,\ldots,N$, each satisfying an ISSf estimate of the form \eqref{tra_ISSf} with inputs formed by specific components of the state $x$ and the external input $w$. In particular, for each $i=1,\ldots,N$, there exist $\beta_i\in\mathcal{K}^-\mathcal{L}$, continuous functions $\sigma_i:\mathcal{D}\to\mathbb{R}$, $\psi_i^x:\mathbb{R}^n\to\mathbb{R}$ and $\psi_i^w:\mathbb{R}^m\to\mathbb{R}$, $\gamma_i,\chi_i\in\mathcal{K}^-$, and a constant $c_i\le 0$ such that, for any $x(0)\in\mathcal{D}$ and any measurable and essentially bounded $w:\mathbb{R}_{+}\to\mathbb{R}^m$, the trajectory $x(t)$ of system \eqref{network_dynamics} satisfies
\begin{align}
h_i(x(t))&\ge \min\{\beta_i(\sigma_i(x(0)),t),\gamma_i(\lfloor\!\lfloor \psi_i^x(x(\cdot))\rfloor\!\rfloor_{[0,t)}),\nonumber\\
&~~~\chi_{i}(\lfloor\!\lfloor \psi_i^w(w(\cdot))\rfloor\!\rfloor_{[0,t)}),c_i\}
\label{ISSf_psi}
\end{align}
for all $t\in[0,t_m)$, where $[0,t_m)$ is the time interval over which $x(t)$ is right-maximally defined. In \eqref{ISSf_psi}, the coupling term $\psi_i^x(x)$ characterizes the influence of specific components of the state $x$ on $h_i(x)$ due to interconnections among the subsystems, while $\psi_i^w(w)$ characterizes the influence of the external input $w$.

Motivated by the example in Section~\ref{section_Motivating_Example} and the existing safety-oriented small-gain results \cite{LYU_2022_auto,Lyu_2024_IEEETAC,WANG_2025_IFAC}, we suppose that the coupling term $\psi_i^x(x)$ admits a lower bound in terms of the other safety indicator functions $h_j(x)$, $j\neq i$. That is, there exist $\tilde{\gamma}_{ij}\in\mathcal{K}^-\cup\{0\}$ for $j\neq i$ such that
\begin{align}
\psi_i^x(x)\ge \min_{\substack{j=1,\ldots,N\\j\neq i}}\tilde{\gamma}_{ij}(h_j(x))
\label{condition_couple_bound}
\end{align}
for all $x\in\mathcal{D}$. Combining conditions \eqref{ISSf_psi} and \eqref{condition_couple_bound}, we make the following assumption.
\begin{assumption}\label{assum_ISSf}
For each $i=1,\ldots,N$, with $h_i:\mathbb{R}^n\to\mathbb{R}$ being a safety indicator function, there exist $\beta_i\in\mathcal{K}^-\mathcal{L}$, $\gamma_{ij}\in\mathcal{K}^-\cup\{0\}$ for $j\neq i$, continuous functions $\sigma_i:\mathcal{D}\to\mathbb{R}$ and $\psi_i^w:\mathbb{R}^m\to\mathbb{R}$, $\chi_i\in\mathcal{K}^-$ and a constant $c_i\le 0$ such that, for any $x(0)\in\mathcal{D}$ and any measurable and essentially bounded $w:\mathbb{R}_{+}\to\mathbb{R}^m$, the trajectory $x(t)$ of system \eqref{network_dynamics} satisfies
\begin{align}
h_i(x(t))&\ge \min\{\beta_i(\sigma_i(x(0)),t),\nonumber\\
&~~~\min_{\substack{j=1,\ldots,N\\ j\neq i}}\gamma_{ij}(\lfloor\!\lfloor h_j(x(\cdot))\rfloor\!\rfloor_{[0,t)}),\nonumber\\
&~~~\chi_{i}(\lfloor\!\lfloor \psi_i^w(w(\cdot))\rfloor\!\rfloor_{[0,t)}),c_i\}
\label{condition_ISSf}
\end{align}
for all $t\in[0,t_m)$, where $[0,t_m)$ is the time interval over which $x(t)$ is right-maximally defined.
\end{assumption}

Property \eqref{condition_ISSf} shows that the safety indicator functions $h_i$ interact with one another, with interactions characterized by the functions $\gamma_{ij}$. We call $\gamma_{ij}$ the gain function, or simply the gain, from $h_j$ to $h_i$.

\begin{assumption}\label{assum_UO}
For each $i=1,\ldots,N$, there exist a function $\alpha_i^o\in\mathcal{K}^-$ and a constant $c^o_i\le 0$ such that for any $x(0)\in\mathcal{D}$ and any measurable and essentially bounded $w:\mathbb{R}_{+}\to\mathbb{R}^m$,
\begin{align}
\sigma_i(x(t))&\ge \alpha_i^o(\min\{\sigma_i(x(0)),\min_{j=1,\ldots,N}\lfloor\!\lfloor h_j(x(\cdot))\rfloor\!\rfloor_{[0,t)},\nonumber\\
&~~~\min_{j=1,\ldots,N}\lfloor\!\lfloor \psi^w_j(w(\cdot))\rfloor\!\rfloor_{[0,t)},c^o_i\})\label{condition_UO}
\end{align}
for all $t\in[0,t_m)$.
\end{assumption}

\begin{remark}
Assumption \ref{assum_UO} is analogous to the unboundedness observability (UO) assumption used in stability-oriented small-gain analysis based on the formulation of input-to-output stability (IOS) \cite{Jiang_1994_MCSS,Jiang_2008_WCICA,BAO_2019_IEEETAC}. Specifically, classical UO typically requires that a system like $\dot{x}=f(x,w)$, $y=g(x,w)$ satisfies
\begin{align}
|x(t)| \le \alpha^s\left(\max\left\{|x(0)|, \|y\|_{[0,t]}, \|w\|_{[0,t]}\right\}\right),\label{classical_UO}
\end{align}
where $x\in\mathbb{R}^n$ is the system state, $y\in\mathbb{R}^p$ is the output, $w\in\mathbb{R}^m$ is the input, $f$ and $g$ are functions ensuring well-posedness of the system, and $\alpha^s\in\mathcal{K}$. Intuitively, UO ensures that the state remains bounded whenever the output is bounded. Note that an ISS system \cite{Sontag1989TAC} by default features the UO property. In safety-oriented small-gain analysis, Assumption \ref{assum_UO} ensures that lower bounds on the safety indicator functions imply lower bounds on the transient effects of the state on these functions, represented by $\sigma_i(x)$, $i=1,\ldots,N$. In the proof of the main result in Section \ref{section_proof_mainresult}, this assumption is used to obtain uniform lower bounds on $\sigma_i(x(\mu t))$ from those on the safety indicator functions, which enables an induction-based small-gain analysis (Lemma \ref{Lemma_convergence_min}) to establish ISSf of the overall system.
\end{remark}

With property \eqref{interconnection_N}, it is straightforward to verify that the controlled multi-agent system defined by \eqref{apply_system_oneorder} and \eqref{apply_ui_definition_N} satisfies Assumptions \ref{assum_ISSf} and \ref{assum_UO}. In particular, for $i,j=1,\ldots,N$, $j\neq i$, one may set $\beta_i=\beta_o$, $\sigma_i=h_i$, $\gamma_{ij}=\hat{\gamma}_{ij}$, $c_i=c_o$, $\chi_i=\chi_o$, $\psi_i^w(w)=-|w_i|$ for $w=[w_1^T,\ldots,w_N^T]^T\in\mathbb{R}^{2N}$ with $w_k\in\mathbb{R}^2$ for $k=1,\ldots,N$, $\alpha_i^o(s)=\min\{s,0\}$ for $s\in\mathbb{R}$, and $c_i^o=0$.

\subsection{Main Results}

Theorem \ref{theorem_smallgain_min} presents a safety-oriented small-gain condition for the overall interconnected system \eqref{network_dynamics}.

\begin{theorem}\label{theorem_smallgain_min}
Consider system \eqref{network_dynamics}. Under Assumptions \ref{assum_ISSf} and \ref{assum_UO}, suppose that the following set of small-gain conditions holds for each simple cycle $(i_1,i_2,\ldots,i_r,i_1)$:
\begin{align}
\gamma_{i_1 i_2} \circ \gamma_{i_2i_3} \circ \cdots \circ \gamma_{i_{r-1}i_{r}}\circ \gamma_{i_ri_1}(s)
&> s, \quad \forall s<0\label{circuitgain}
\end{align}
for $r =2,\ldots,N$ and $i_j \in \{1,\ldots,N\}$, $i_j\neq i_{j'}$ if $j\neq j'$. Then, for any $x(0)\in\mathcal{D}$ and any measurable and essentially bounded $w:\mathbb{R}_{+}\to\mathbb{R}^m$,
\begin{align}
\mathbf{H}(t)\ge \operatorname{MIN}\{Q(\Omega(x(0))),Q(\mathbf{w}(t)),Q(\mathbf{c})\}\label{h_lowerbound_min_theorem}
\end{align}
holds for all $t\in[0,t_m)$, where
\begin{subequations}
\begin{align}
\Omega(\xi)&=(\beta_1(\sigma_1(\xi),0),\ldots,\beta_N(\sigma_N(\xi),0))\label{definition_Omega}\\
Q(\mathbf{s}) &= \operatorname{MIN}\{
\mathbf{s}, \Gamma(\mathbf{s}), \ldots, [\Gamma]^{N-1}(\mathbf{s})\},\label{definition_Q_theorem}\\
\Gamma(\mathbf{s}) &= (\Gamma_{1}(\mathbf{s}), \ldots, \Gamma_{N}(\mathbf{s})),\label{definition_Gamma_min}\\
\Gamma_{i}(\mathbf{s})&=\min_{\substack{j=1,\ldots,N\\j\neq i}}\gamma_{ij}(s_j), \quad i=1,\ldots,N\label{definition_Gamma_i_min}
\end{align}
\end{subequations}
for $\xi\in\mathcal{D}$ and $\mathbf{s}=[s_1,\ldots,s_N]^T\in \mathbb{R}^{N}$, and
\begin{subequations}
\begin{align}
\mathbf{H}(t)&=(h_1(x(t)),\ldots,h_N(x(t)))\label{definition_H},\\
\mathbf{w}(t)&=(\chi_{1}(\lfloor\!\lfloor \psi^w_1(w(\cdot))\rfloor\!\rfloor_{[0,t)}),\ldots,\chi_{N}(\lfloor\!\lfloor \psi^w_N(w(\cdot))\rfloor\!\rfloor_{[0,t)}))\label{definition_d},\\
\mathbf{c}&=(c_1,\ldots,c_N)\label{definition_c}.
\end{align}
\end{subequations}
Moreover,
\begin{align}
h(x(t))\ge \min\{\beta(\sigma(x(0)),t),\chi(\lfloor\!\lfloor \psi^w(w(\cdot))\rfloor\!\rfloor_{[0,t)}),c\}\label{interconnected_ISSf}
\end{align}
for all $t\in[0,t_m)$, where
\begin{align}
h(x)=\min_{i=1,\ldots,N}h_i(x)\label{definition_h}
\end{align}
for $x\in\mathcal{D}$, $\beta\in \mathcal{K}^-\mathcal{L}$, $\chi\in\mathcal{K}^-$, $c\le 0$ is a constant, $\psi^w:\mathbb{R}^m\to\mathbb{R}$ and $\sigma:\mathcal{D}\to\mathbb{R}$ are continuous functions satisfying
\begin{align}
\sigma(x(t))&\ge \alpha^o(\min\{\sigma(x(0)),\min_{i=1,\ldots,N}\lfloor\!\lfloor h_i(x(\cdot))\rfloor\!\rfloor_{[0,t)},\nonumber\\
&~~~\min_{i=1,\ldots,N}\lfloor\!\lfloor \psi^w_i(w(\cdot))\rfloor\!\rfloor_{[0,t)},\min_{j=1,\ldots,N}c_j^o\})\label{definition_sigma_UO}
\end{align}
for $t\in[0,t_m)$ with $\alpha^o\in\mathcal{K}^-$.
\end{theorem}

The proof of Theorem \ref{theorem_smallgain_min} is given in Section \ref{section_proof_mainresult}.

Corollary \ref{corollary_sigma} considers the special case where $\sigma_i(x)=h_i(x)$ for all $i=1,\ldots,N$.

\begin{corollary}\label{corollary_sigma}
Consider system \eqref{network_dynamics}. Under Assumption \ref{assum_ISSf}, if for each $i=1,\ldots,N$,
\begin{align}
\sigma_i(x)=h_i(x), \quad i=1,\ldots,N\label{sigma_h_condition}
\end{align}
for all $x\in\mathcal{D}$, and the small-gain condition \eqref{circuitgain} holds, then properties \eqref{h_lowerbound_min_theorem} and \eqref{interconnected_ISSf} hold.
\end{corollary}

It is common in formation control \cite{Tanner_2004_TRO} and vehicular platooning \cite{Ploeg_2014_TCST,Liu_2023_TIV} scenarios that the interconnected system features a cascade structure or a necklace structure, which can be regarded as special cases of the network structure considered in Theorem \ref{theorem_smallgain_min}. 

\begin{corollary}\label{corollary_cascade}
Consider system \eqref{network_dynamics}. Under Assumptions \ref{assum_ISSf} and \ref{assum_UO}, if
\begin{align}
\gamma_{ij}(s)=0, \quad \forall i=1,\ldots,N, \;j>i\label{cascade_small_gain_condition}
\end{align}
for all $s\in\mathbb{R}$, then properties \eqref{h_lowerbound_min_theorem} and \eqref{interconnected_ISSf} hold.
\end{corollary}

\begin{corollary}\label{corollary_one_loop}
Consider system \eqref{network_dynamics}. Under Assumptions \ref{assum_ISSf} and \ref{assum_UO}, if
\begin{align}
\gamma_{12}\circ\gamma_{23}\circ\cdots\circ\gamma_{(N-1)N}\circ\gamma_{N1}(s)>s,\quad \forall s<0\label{one_loop_small_gain_condition_1}
\end{align}
and all the other gain functions satisfy
\begin{align}
\gamma_{ij}(s)&=0,\quad \forall s\in\mathbb{R},\nonumber\\
(i,j)&\notin\{(1,2),(2,3),\ldots,(N-1,N),(N,1)\},\label{one_loop_small_gain_condition_2}
\end{align}
then properties \eqref{h_lowerbound_min_theorem} and \eqref{interconnected_ISSf} hold.
\end{corollary}

Corollary \ref{corollary_subsystem} considers the large-scale systems in which each subsystem corresponds to one safety indicator function.

\begin{corollary}\label{corollary_subsystem}
Theorem \ref{theorem_smallgain_min} remains valid when the safety indicator functions are in the form of
\begin{equation}
h_i(x)=\bar{h}_i(x_i)
\end{equation}
for $i=1,\ldots,N$, where each $x_i$ is a portion of $x$ such that
$x=[x_1^T,\ldots,x_N^T]^T$.
\end{corollary}

\begin{remark}
Corresponding to the setup in Corollary \ref{corollary_subsystem}, in \cite{LYU_2022_auto,Lyu_2024_IEEETAC,WANG_2025_IFAC}, the functions used to characterize safety properties of the subsystems are assumed to satisfy gradient-based conditions, with one such function associated with each subsystem; that is, for each $i=1,\ldots,N$, there exist $\hat{\gamma}_{ij}\in\mathcal{K}^e\cup\{0\}$ for $j\neq i$ and $\chi_i,\alpha_i\in\mathcal{K}^e$ such that
\begin{align}
&\bar{h}_{i}(x_i)\le \min_{j\neq i}\{\hat{\gamma}_{ij}(\bar{h}_j(x_j)),\chi_i(-|w|)\}\nonumber\\
\Rightarrow&D^+\bar{h}_i(x_i;f_i(x,w))\ge -\alpha_i(\bar{h}_i(x_i))\label{otherpaper_assum}
\end{align}
for all $x=[x_1^T,\ldots,x_N^T]^T\in\mathbb{R}^n$, where for each $i=1,\ldots,N$, $\bar{h}_i:\mathbb{R}^{n_i}\to\mathbb{R}$ is locally Lipschitz continuous, $f_i(x,w)$ denotes the component of the vector field $f(x,w)$ in \eqref{network_dynamics} corresponding to $x_i$, i.e., $\dot{x}_i=f_i(x,w)$. It should be noted that the $-|w|$ term in \eqref{otherpaper_assum} is replaced by $\lfloor w\rfloor$ in the formulation in \cite{WANG_2025_IFAC}. By Theorem \ref{ISSf-BFtheorem}, if system \eqref{network_dynamics} satisfies condition \eqref{otherpaper_assum}, then Assumptions \ref{assum_ISSf} and \ref{assum_UO} hold with $h_i(x)=\bar{h}_i(x_i)$, $\gamma_{ij}(s)=\min\{\hat{\gamma}_{ij}(s),0\}$, $\alpha_i^o(s)=\min\{s,0\}$, $\sigma_i(x)=\bar{h}_i(x_i)$, $\psi_i^w(w)=-|w|$, and $c_i^o=0$,
for all $s\in\mathbb{R}$,
$x=[x_1^T,\ldots,x_N^T]^T\in\mathbb{R}^n$,
and $w\in\mathbb{R}^m$. Thus, if small-gain condition \eqref{circuitgain} is satisfied, then the safety of system \eqref{network_dynamics} can be verified using Corollary \ref{corollary_subsystem}. Based on property \eqref{otherpaper_assum}, the small-gain condition in \cite{LYU_2022_auto,Lyu_2024_IEEETAC} requires that the magnitude of the loop gain function be smaller than that of the identity function on both the positive and negative real axes, i.e., for every simple cycle $(i_1,i_2,\ldots,i_r,i_1)$,
\begin{align}
|\hat{\gamma}_{i_1 i_2} \circ \hat{\gamma}_{i_2i_3} \circ \cdots \circ \hat{\gamma}_{i_{r-1}i_{r}}\circ \hat{\gamma}_{i_ri_1}(s)|
&< |s|, \quad \forall s\in\mathbb R\setminus\{0\},\label{circuitgain_otherpaper}
\end{align}
where $r =2,\ldots,N$ and $i_j \in \{1,\ldots,N\}$, $i_j\neq i_{j'}$ if $j\neq j'$. This requirement is stronger than our small-gain condition \eqref{circuitgain}. Meanwhile, the small-gain condition in \cite{WANG_2025_IFAC} is consistent
with the form of condition \eqref{circuitgain}, but relies on the gradient-based
formulation \eqref{otherpaper_assum}.
\end{remark}

\begin{remark}\label{remark_ISSfequivalenceRobustSafety}
In the context of stability analysis, a system is ISS if and only if it is robustly stable \cite[Theorem 1]{SONTAG_1995_SCL}. Such an equivalence provides a powerful framework for understanding robust stability, and for applying tools from interconnected systems to robustness analysis. However, to the best of our knowledge, such a result has not been fully investigated in the existing literature on safety-critical control. It is of interest to note that the safety-oriented small-gain theorem in this paper naturally establishes a relationship between ISSf and robust safety; see, e.g., \cite[Proposition 5]{Xu2015IFAC} for the notion of robust safety. Consider system \eqref{network_dynamics} with
\begin{align}
w=\kappa(t,x)\label{disturbance_k}
\end{align}
where $\kappa:\mathbb{R}_+\times\mathbb{R}^n\to\mathbb{R}^m$ is locally Lipschitz continuous with respect to $x$ on $\mathcal{D}$, and continuous with respect to $t$. Suppose that system \eqref{network_dynamics} is ISSf with respect to some continuous $h$ in the sense of \eqref{tra_ISSf}, and that $\sigma(x)=h(x)$, $\psi(w)=\lfloor w \rfloor$, and $c=0$ in \eqref{tra_ISSf}. With Theorem \ref{theorem_smallgain_min}, along the trajectories of the interconnected system composed of \eqref{network_dynamics} and \eqref{disturbance_k}, it holds that
\begin{align}
h(x(t))\ge \hat{\beta}^r(h(x(0)),t)
\end{align}
with $\hat{\beta}^r\in\mathcal{K}^-\mathcal{L}$, whenever $x(t)$ is defined, provided that the function $\kappa$ satisfies 
\begin{align}
\lfloor \kappa(t,x) \rfloor\ge \rho(h(x))
\end{align}
for all $t\ge 0$ and all $x\in\mathcal{D}$, with $\rho\in\mathcal{K}^-$ satisfying the small-gain condition
\begin{align}
\gamma\circ\rho(s)>s,\quad\forall s<0.
\end{align}
\end{remark}

\section{Proof of the Main Result}\label{section_proof_mainresult}

In the proof, we treat the interconnected system \eqref{network_dynamics} as a vector system with a vector-valued self-loop described by $\Gamma$, defined in \eqref{definition_Gamma_min}. Under the small-gain condition \eqref{circuitgain}, we first establish lower boundedness of the safety indicator functions. Then, we show that the self-loop can be removed by using a $\mathcal{K}^-\mathcal{L}$ function $\hat{\beta}$ smaller than $\beta_i$ instead to characterize the decay rate of the overall system. A key technique used in the proof is an inductive comparison argument that constructs successive lower bounds along a sequence of time instants and uses an auxiliary signal vector defined on $[0,\infty)$ to obtain $\mathcal{K}^-\mathcal{L}$ estimates throughout the maximal existence interval, which is presented as a technical lemma in Appendix \ref{appendix_lemma_convergence}.

The following discussions apply to any trajectory of the overall system with initial state $x(0)\in\mathcal{D}$ and the external input $w:\mathbb{R}_{+}\to\mathbb{R}^m$ being measurable and essentially bounded. The time interval right-maximally defining the Carath\'eodory solution $x(t)$ is denoted by $[0,t_m)$.

\subsection{Lower Boundedness of $h(x(t))$}

Denote
\begin{align}
\mathbf{H}_m(t)&=(\lfloor\!\lfloor h_1(x(\cdot))\rfloor\!\rfloor_{[0,t)},\ldots,\lfloor\!\lfloor h_N(x(\cdot))\rfloor\!\rfloor_{[0,t)}).\label{definition_H_m}
\end{align}
For $i=1,\ldots,N$, with the definition of $\Gamma_i$ in \eqref{definition_Gamma_i_min}, property \eqref{condition_ISSf} directly implies that
\begin{align}
\lfloor\!\lfloor h_i(x(\cdot))\rfloor\!\rfloor_{[0,t)}&\ge \min\{\beta_i(\sigma_i(x(0)),0),\Gamma_{i}(\mathbf{H}_m(t)),\nonumber\\
&~~~\chi_{i}(\lfloor\!\lfloor \psi_i^w(w(\cdot))\rfloor\!\rfloor_{[0,t)}),c_i\}
\end{align}
for all $t\in[0,t_m)$. Thus,
\begin{align}
\mathbf{H}_m(t)\ge \operatorname{MIN}\{\Omega(x(0)),\Gamma(\mathbf{H}_m(t)),\mathbf{w}(t),\mathbf{c}\}\label{vector_subsystemISSf}
\end{align}
for all $t\in[0,t_m)$, where $\Omega$ is defined in \eqref{definition_Omega}, $\mathbf{w}$ is defined in \eqref{definition_d}, and $\mathbf{c}$ is defined in \eqref{definition_c}.

Since $\gamma_{ij}\in\mathcal{K}^-\cup\{0\}$ for $i,j=1,\ldots,N$, $j\neq i$, it holds that $\Gamma$ defined in \eqref{definition_Gamma_min} and \eqref{definition_Gamma_i_min} with the $\min$ operation admits the $\operatorname{MIN}$-preserving property, that is,
\begin{subequations}
\begin{align}
&\mathbf{s} \ge \mathbf{s}' \Rightarrow \Gamma(\mathbf{s}) \ge \Gamma(\mathbf{s}')\label{min_preserving1}\\
&\Gamma(\operatorname{MIN}\{\mathbf{s},\mathbf{s}'\})
= \operatorname{MIN}\{\Gamma(\mathbf{s}),\Gamma(\mathbf{s}')\}\label{min_preserving2}
\end{align}
\end{subequations}
for all $\mathbf{s},\mathbf{s}'\in\mathbb{R}^N$.

By substituting \eqref{vector_subsystemISSf} into itself and using properties \eqref{min_preserving1} and \eqref{min_preserving2}, we have
\begin{align}
\mathbf{H}_m(t)&\ge\operatorname{MIN}\{\Omega(x(0)),\mathbf{w}(t),\mathbf{c},\nonumber\\
&~~~\Gamma(\operatorname{MIN}\{\Omega(x(0)),\Gamma(\mathbf{H}_m(t)),\mathbf{w}(t),\mathbf{c}\})\}\nonumber\\
&\ge\operatorname{MIN}\{
\Omega(x(0)), \Gamma(\Omega(x(0))), \mathbf{w}(t),
\Gamma(\mathbf{w}(t)), \nonumber\\
&~~~[\Gamma]^{2}(\mathbf{H}_m(t)),\mathbf{c},\Gamma(\mathbf{c})\}.
\end{align}

Repeating this substitution $(q^*-1)$ times yields
\begin{align}
\mathbf{H}_m(t)
&\ge \operatorname{MIN}\{\Omega(x(0)), \Gamma(\Omega(x(0))),\ldots,[\Gamma]^{q^*-1}(\Omega(x(0))),\nonumber\\
&~~~[\Gamma]^{q^*}(\mathbf{H}_m(t)),\mathbf{w}(t),\Gamma(\mathbf{w}(t)),\ldots,[\Gamma]^{q^*-1}(\mathbf{w}(t)),\nonumber\\
&~~~\mathbf{c},\Gamma(\mathbf{c}),\ldots,[\Gamma]^{q^*-1}(\mathbf{c})\}.\label{H_m_q*}
\end{align}
Define
\begin{align}
\bar{\Gamma}(\mathbf{s})&=-\Gamma(-\mathbf{s}),\quad
\bar{Q}(\mathbf{s})=-Q(-\mathbf{s}),\label{definition_Qbar}\\
\bar{\gamma}_{ij}(s)&=-\gamma_{ij}(-s)
\end{align}
for $\mathbf{s}\in\mathbb{R}^N$ and $s\in\mathbb{R}$, where $Q$ is defined in \eqref{definition_Q_theorem}. Under condition \eqref{circuitgain}, it holds that for every simple cycle $(i_1,i_2,\ldots,i_r,i_1)$,
\begin{align}
\bar{\gamma}_{i_1 i_2} \circ \bar{\gamma}_{i_2i_3} \circ \cdots \circ \bar{\gamma}_{i_{r-1}i_{r}}\circ \bar{\gamma}_{i_ri_1}(s)
&< s, \quad \forall s>0\label{circuitgain_bar}
\end{align}
where $r =2,\ldots,N$ and $i_j \in \{1,\ldots,N\}$, $i_j\neq i_{j'}$ if $j\neq j'$. Then, by \cite[Lemma 2.3.14]{RUFFER_2007_phd},
\begin{align}
\bar{\Gamma}(-\mathbf{s})\ngeq -\mathbf{s}\label{smallgain_condition_Gammabar_theorem}
\end{align}
for all $\mathbf{s}< \mathbf{0}_N$. By using property \eqref{smallgain_condition_Gammabar_theorem}, with \cite[Proposition 2.7]{karafyllis_2011_vector}, for $k\in\mathbb{Z}_+$,
\begin{align}
[\bar{\Gamma}]^k(-\mathbf{s})\le \bar{Q}(-\mathbf{s}),\quad -[\bar{\Gamma}]^k(-\mathbf{s})\ge -\bar{Q}(-\mathbf{s})
\end{align}
for all $\mathbf{s}\le \mathbf{0}_N$. With the definitions of $\bar{\Gamma}$ and $\bar{Q}$ in \eqref{definition_Qbar}, for each $k\in\mathbb{Z}_+$, we have
\begin{align}
[\Gamma]^k(\mathbf{s})\ge Q(\mathbf{s})\label{Gamma_k_lowerbound}
\end{align}
for all $\mathbf{s}\le \mathbf{0}_N$. Substituting property \eqref{Gamma_k_lowerbound} into \eqref{H_m_q*}, we have
\begin{align}
\mathbf{H}_m(t) 
&\ge \operatorname{MIN}\{
Q(\Omega(x(0))),[\Gamma]^{q^{*}}(\mathbf{H}_m(t)),Q(\mathbf{w}(t)),Q(\mathbf{c})\}\label{H_Gamma_q}
\end{align}
for all $t\in[0,t_m)$ and every positive integer $q^*$. Under condition \eqref{smallgain_condition_Gammabar_theorem}, by \cite[Theorem 2.2.8]{RUFFER_2007_phd}, it holds that $\lim_{k\to\infty}[\bar{\Gamma}]^k(-\mathbf{s})=\mathbf{0}_N $ for all $\mathbf{s}\in\mathbb{R}^N$. Then, we have
\begin{align}
-\lim_{k\to\infty}[\Gamma]^k(\mathbf{s})=\lim_{k\to\infty}[\bar{\Gamma}]^k(-\mathbf{s})=\mathbf{0}_N\label{Gammabar_k_convergence}
\end{align}
for all $\mathbf{s}\le \mathbf{0}_N$. Therefore,
\begin{align}
\lim_{k\to\infty}[\Gamma]^k(\mathbf{s})\ge \lim_{k\to\infty}[\Gamma]^k(\operatorname{MIN}\{\mathbf{s},\mathbf{0}_N\})=\mathbf{0}_N\label{Gamma_k_convergence}
\end{align}
for all $\mathbf{s}\in\mathbb{R}^N$. Fix any $t\in[0,t_m)$. By the continuity of the safety
indicator functions, the vector
$\mathbf{H}_m(t)$ is finite. Since
$\Gamma(\mathbf{s})\le\mathbf{0}_N$ for all
$\mathbf{s}\in\mathbb{R}^N$, property
\eqref{Gamma_k_convergence} implies that
\begin{align}
\lim_{q\to\infty}[\Gamma]^q(\mathbf{H}_m(t))
=\mathbf{0}_N.
\end{align}
Combining properties \eqref{H_Gamma_q} and \eqref{Gamma_k_convergence}, we conclude that, for any $\epsilon<0$, there exists a $q^*$ such that
\begin{align}
\mathbf{H}_m(t)
&\ge \operatorname{MIN}\{
Q(\Omega(x(0))),[\Gamma]^{q^{*}}(\mathbf{H}_m(t)),Q(\mathbf{w}(t)),Q(\mathbf{c})\}\nonumber\\
&\ge \operatorname{MIN}\{Q(\Omega(x(0))),\epsilon\mathbf{1}_N,Q(\mathbf{w}(t)),Q(\mathbf{c})\}
\label{lowerboundness_h}.
\end{align}
Thus, we have
\begin{align}
\mathbf{H}_m(t)
&\ge \operatorname{MIN}\{
Q(\Omega(x(0))),\mathbf{0}_N,Q(\mathbf{w}(t)),Q(\mathbf{c})\}\nonumber\\
&= \operatorname{MIN}\{
Q(\Omega(x(0))),Q(\mathbf{w}(t)),Q(\mathbf{c})\}\label{H_m_lower_bound}
\end{align}
where the equality follows from
$\operatorname{MIN}\{\Omega(x(0)),\mathbf{w}(t),\mathbf{c}\}
\le \mathbf{0}_N$ and $Q(\mathbf{s})\le\mathbf{s}$. With the definition of $\mathbf{H}_m(t)$ in \eqref{definition_H_m}, property \eqref{h_lowerbound_min_theorem} is established.

\subsection{ISSf of the Overall Interconnected System}

Due to time invariance and causality, for each $i=1,\ldots,N$, property \eqref{condition_ISSf} implies
\begin{align}
h_i(x(t))&\ge \min\{\beta_i(\sigma_i(x(\mu t)),(1-\mu)t),\nonumber\\
&~~~\min_{j\neq i}\gamma_{ij}(\lfloor\!\lfloor h_j(x(\cdot))\rfloor\!\rfloor_{[\mu t,t)}),\nonumber\\
&~~~\chi_{i}(\lfloor\!\lfloor \psi_i^w(w(\cdot))\rfloor\!\rfloor_{[0,\infty)}),c_i\}\label{hi_mut}
\end{align}
for all $t\in[0,t_m)$, where the constant $\mu$ is arbitrarily chosen from $(0,1)$. 

Under condition \eqref{condition_UO}, by using property \eqref{h_lowerbound_min_theorem}, we have
\begin{align}
\sigma_i(x(\mu t))&\ge \alpha_i^o(\min\{\sigma_i(x(0)),\min_{j=1,\ldots,N}\lfloor\!\lfloor h_j(x(\cdot))\rfloor\!\rfloor_{[0,\mu t)},\nonumber\\
&~~~\min_{j=1,\ldots,N}\lfloor\!\lfloor \psi_j^w(w(\cdot))\rfloor\!\rfloor_{[0,\infty)},c_i^o\})\nonumber\\
&\ge \alpha_i^o(\min\{\sigma_i(x(0)),\lfloor Q(\Omega(x(0)))\rfloor,\lfloor Q(\mathbf{w}_\infty)\rfloor,\nonumber\\
&~~~\lfloor Q(\mathbf{c})\rfloor,\min_{j=1,\ldots,N}\lfloor\!\lfloor \psi_j^w(w(\cdot))\rfloor\!\rfloor_{[0,\infty)},c_i^o\})\label{sigma_underline}
\end{align}
for $t\in[0,t_m)$, where
\begin{align}
\mathbf{w}_\infty=(\chi_{1}(\lfloor\!\lfloor \psi_1^w(w(\cdot))\rfloor\!\rfloor_{[0,\infty)}),\ldots,\chi_{N}(\lfloor\!\lfloor \psi_N^w(w(\cdot))\rfloor\!\rfloor_{[0,\infty)}))\label{definition_w_infty}
\end{align} 
Denote
\begin{align}
\underline{h}&=\min_{i=1,\ldots,N}\alpha_i^o(\min\{\sigma_i(x(0)),\lfloor Q(\Omega(x(0)))\rfloor,\lfloor Q(\mathbf{w}_\infty)\rfloor,\nonumber\\
&~~~\lfloor Q(\mathbf{c})\rfloor,\min_{j=1,\ldots,N}\lfloor\!\lfloor \psi_j^w(w(\cdot))\rfloor\!\rfloor_{[0,\infty)},c_i^o\})\label{definition_h_underline}.
\end{align}
Substituting \eqref{sigma_underline} and \eqref{definition_h_underline} into \eqref{hi_mut} gives
\begin{align}
h_i(x(t))&\ge \min\{\beta_i(\underline{h},(1-\mu )t),\nonumber\\
&~~~\min_{j\neq i}\gamma_{ij}(\lfloor\!\lfloor h_j(x(\cdot))\rfloor\!\rfloor_{[\mu t,t)}),\nonumber\\
&~~~\chi_{i}(\lfloor\!\lfloor \psi_i^w(w(\cdot))\rfloor\!\rfloor_{[0,\infty)}),c_i\}.\label{hi_mut_underline}
\end{align}
With the definitions of $\Gamma$ in \eqref{definition_Gamma_min} and \eqref{definition_Gamma_i_min}, $\mathbf{w}_\infty$ in \eqref{definition_w_infty}, $\mathbf{c}$ in \eqref{definition_c} and $\mathbf{H}$ in \eqref{definition_H}, property \eqref{hi_mut_underline} can be rewritten as
\begin{align}
\mathbf{H}(t)\ge \operatorname{MIN}\{\mathbf{B}(\underline{h},t),\Gamma(\mathbf{H}_{[\mu t,t)}),\mathbf{w}_\infty,\mathbf{c}\},\label{self_loop_min}
\end{align}
where 
\begin{align}
\mathbf{B}(s,t)=(\beta_1(s,(1-\mu)t),\ldots,\beta_N(s,(1-\mu)t))
\end{align}
for $s\in\mathbb{R}$ and $t\ge 0$, and
\begin{align}
\mathbf{H}_{[\mu t,t)}=(\lfloor\!\lfloor h_1(x(\cdot))\rfloor\!\rfloor_{[\mu t,t)},\ldots,\lfloor\!\lfloor h_N(x(\cdot))\rfloor\!\rfloor_{[\mu t,t)})
\end{align}
for $t\in[0,t_m)$.

Based on property \eqref{self_loop_min}, with Lemma \ref{Lemma_convergence_min}, we can verify the existence of $\hat{\beta}\in\mathcal{K}^-\mathcal{L}$ such that
\begin{align}
\lfloor\mathbf{H}(t)\rfloor&\ge \min\{\hat{\beta}(\underline{h},t),\lfloor Q_m(\operatorname{MIN}\{\mathbf{w}_\infty,\mathbf{c}\})\rfloor\}\nonumber\\
&\ge \min\{\hat{\beta}(\underline{h},t),\lfloor Q_m(\mathbf{w}_\infty)\rfloor,\lfloor Q_m(\mathbf{c})\rfloor\}\label{Hrfloor_convergence_}
\end{align}
for all $t\in[0,t_m)$, where
\begin{align}
Q_m(\mathbf{s})=\operatorname{MIN}\{\mathbf{s},\Gamma_m(\mathbf{s}),[\Gamma_m]^2(\mathbf{s}),\ldots,[\Gamma_m]^{N-1}(\mathbf{s})\}
\end{align}
for $\mathbf{s}\in\mathbb{R}^N$ with 
\begin{align}
\Gamma_m(\mathbf{s})=\operatorname{MIN}\{\Gamma(\mathbf{s}),\lambda\mathbf{s}\}
\end{align}
and a constant $\lambda\in(0,1)$.
From the definition of $h$ in \eqref{definition_h}, we have
\begin{align}
h(x(t))=\lfloor\mathbf{H}(t)\rfloor.
\end{align}
Then, based on property \eqref{Hrfloor_convergence_} with $\underline{h}$ defined in \eqref{definition_h_underline}, we have
\begin{align}
h(x(t))&\ge \min\{\hat{\beta}(\underline{h},t),\lfloor Q_m(\mathbf{w}_\infty)\rfloor,\lfloor Q_m(\mathbf{c})\rfloor\}\nonumber\\
&\ge \min\{\hat{\beta}(\min_{i=1,\ldots,N}\alpha_i^o(\min\{\sigma_i(x(0)),\nonumber\\
&~~~\lfloor Q(\Omega(x(0)))\rfloor,c_i^o\}),t),\nonumber\\
&~~~\hat{\beta}(\min_{i=1,\ldots,N}\alpha^o_i(\lfloor Q(\mathbf{w}_\infty)\rfloor),0),\nonumber\\
&~~~\hat{\beta}(\min_{i=1,\ldots,N}\alpha^o_i(\min_{j=1,\ldots,N}\lfloor\!\lfloor \psi_j^w(w)\rfloor\!\rfloor_{[0,\infty)}),0),\nonumber\\
&~~~\hat{\beta}(\min_{i=1,\ldots,N}\alpha^o_i(\lfloor Q(\mathbf{c})\rfloor),0),\lfloor Q_m(\mathbf{w}_\infty)\rfloor,\lfloor Q_m(\mathbf{c})\rfloor\}
\end{align}
for all $t\in[0,t_m)$. Recall the definitions of $\mathbf{w}$ in \eqref{definition_d} and $\mathbf{c}$ in \eqref{definition_c}. Finally, by causality, we can prove \eqref{interconnected_ISSf} by defining
\begin{subequations}\label{definition_sigma_conclude}
\begin{align}
\beta(s,t)&=\hat{\beta}(s,t),\\
\sigma(\xi)&=\min_{i=1,\ldots,N}\alpha_i^o(\min\{\sigma_i(\xi),\lfloor Q(\Omega(\xi))\rfloor,c_i^o\})\label{sigma_large}\\
\psi^w(\zeta)&=\min_{i=1,\ldots,N}\psi^w_i(\zeta)\\
\chi(s)&=\min\{\hat{\beta}(\min_{i=1,\ldots,N}\alpha_i^o(\lfloor Q((\chi_1(s),\ldots,\chi_N(s)))\rfloor),0),\nonumber\\
&~~~\hat{\beta}(\min_{i=1,\ldots,N}\alpha_i^o(s),0),\nonumber\\
&~~~\lfloor Q_m((\chi_1(s),\ldots,\chi_N(s)))\rfloor,s\}\\
c&=\min\{\hat{\beta}(\min_{i=1,\ldots,N}\alpha^o_i(\lfloor Q(\mathbf{c})\rfloor),0),\lfloor Q_m(\mathbf{c})\rfloor\}
\end{align}
\end{subequations}
for $\xi\in\mathcal{D}$, $\zeta\in\mathbb{R}^m$, $s\in\mathbb{R}$, and $t\ge 0$. Moreover, by the definition of $\sigma$ in \eqref{sigma_large} and condition \eqref{condition_UO}, we have \eqref{definition_sigma_UO}.

\begin{remark}\label{remark_introduce_proof}
In the proof of Theorem \ref{theorem_smallgain_min}, we extend the proof in \cite{Jiang_1994_MCSS} based on mathematical induction to establish the ISSf property for the overall system. Specifically, by using boundedness and contractivity properties of the vector-valued gain functions \cite{dashkovskiy_2007_MCSS,karafyllis_2011_vector}, namely, \eqref{Gamma_k_lowerbound} and \eqref{Gamma_k_convergence}, we show the lower boundedness of $h(x(t))$ along the trajectories of the overall system. In the convergence analysis, we define an auxiliary signal vector $\mathbf{v}:\mathbb{R}_+\to\mathbb{R}^N$ and construct a time sequence $\{\bar{t}_k\}_{k\in\mathbb{Z}_+}$. By extending the inductive analysis in \cite{Jiang_1994_MCSS}, we prove the existence of an auxiliary signal vector $\mathbf{v}$ satisfying $\mathbf{v}(t)\le \mathbf{H}(t)$ for $t\in[0,t_m)$ and $\mathbf{v}(t)\ge\operatorname{MIN}\{[\Gamma_m]^k(Q_m(\mathbf{B}(\underline{h},0))),Q_m(\mathbf{w}_\infty),Q_m(\mathbf{c})\}$ for $t\in[\bar{t}_k,\infty)$. The definition of the auxiliary signal, the construction of the discrete time sequence, and the inductive argument are provided in the proof of Lemma \ref{Lemma_convergence_min}.
\end{remark}

\begin{remark}\label{remark_difficult}
It is worth noting that the analysis in \cite{dashkovskiy_2007_MCSS} also offers a potential route to establishing the safety-oriented small-gain theorem. A possible first step is to construct an auxiliary signal vector $\mathbf{z}:\mathbb{R}_+\to\mathbb{R}^N$ satisfying $\mathbf{z}(t)\leq\mathbf{H}(t)$ for all $t\in[0,t_m)$ and the self-loop inequality \eqref{lemma_z_condition_min} for all $t\geq 0$, with $\xi$ in \eqref{lemma_z_condition_min} replaced by $\lfloor\mathbf{z}(0)\rfloor$. Then, under the small-gain condition \eqref{circuitgain}, one may consider establishing a global lower bound for $\mathbf{z}(t)$ depending on $\mathbf{z}(0)$ and a uniform asymptotic lower bound independent of $\mathbf{z}(0)$. Analogously to the characterization of ISS in terms of the asymptotic gain property and zero-input global stability, combining the lower bounds could yield an ISSf-type estimate for $\mathbf{z}(t)$ and thereby establish ISSf of the overall system by using the comparison principle. However, constructing an auxiliary signal vector $\mathbf{z}$ is nontrivial: $\mathbf{z}$ must be defined on the entire nonnegative time axis, satisfy $\mathbf{z}(t)\leq\mathbf{H}(t)$ for all $t\in[0,t_m)$, and obey the self-loop inequality \eqref{lemma_z_condition_min} for all $t\geq0$, even when $t_m$ is finite.
\end{remark}

\section{An Application to Safety-Critical Control of Multi-Agent Systems Subject to External Disturbances}\label{section_apply}

According to Proposition \ref{prop_robust_safety_controller}, the controlled multi-agent system defined by \eqref{apply_system_oneorder} and \eqref{apply_ui_definition_N} satisfies Assumptions \ref{assum_ISSf} and \ref{assum_UO}. Then, we can establish the ISSf property of the controlled multi-agent system and verify that the imposed safety constraints are satisfied.

\subsection{Safety Verification of the Half-Spring Controller}

Our main result on the safety of multi-agent systems with the ``half-spring'' controller is presented in Theorem \ref{theorem_example_uncertain}.

\begin{theorem}\label{theorem_example_uncertain}
Consider the multi-agent system \eqref{apply_system_oneorder}, the obstacles located at $o_1,\ldots,o_{N_o}$, and the controller \eqref{apply_ui_definition_N}. Suppose that the positions of the obstacles satisfy
\begin{align}
\min_{i\neq j}\{|o_i-o_j|\} &> 2d_o+2(N+1) d_a,\label{condition_distance_un}
\end{align}
and there exist nonnegative constants $\bar{v}_c$ and $\bar{w}$ such that \eqref{v_c_condition} and 
\begin{align}
\|w_i\|_{[0,\infty)}\le \bar{w},\quad i=1,\ldots,N\label{d_condition}
\end{align}
hold. Choose the controller parameters in \eqref{apply_ui_definition_N_equivalent} to satisfy
\begin{subequations}\label{parameter_choose_N}
\begin{align}
\lambda&\in(\frac{1-(N+1)\tilde{\lambda}}{N+1},\frac{1}{N+1}),\label{c_choose_un}\\
k_s&>  \max\{\bar{v}_c,\bar{w}\}\max\left\{\frac{1}{\lambda d_{ma}},\frac{4}{2\lambda d_{ma}-(1+N\lambda)d_a}\right\},\label{k_s_choose_un}\\
d_{mo}&=d_o+d_{ma},\label{d_mo_choose_un}\\
d_{ma}&=d_a+\frac{N+(N+1)\tilde{\lambda}}{1-(N+1)\tilde{\lambda}}d_a,
\label{d_ma_choose_un}
\end{align}
\end{subequations}
where
\begin{align}
\tilde{\lambda}&=\frac{\mu d_m}{2 (N+1)^2 d_a+\mu (N+1)d_m}\label{c_tilde_N}
\end{align}
with a constant $\mu\in(0,1)$ and $d_m=\min_{i\neq j}\{|o_i-o_j|\}-2d_o-2(N+1) d_a$. 

Then, there exist $\beta\in\mathcal{K}^-\mathcal{L}$, $\chi\in\mathcal{K}^-$, and a constant $c\le 0$ such that for any $p(0)\in\mathcal{D}_s$, any measurable $v_i^c:\mathbb{R}_{+}\to\mathbb{R}^2$ satisfying \eqref{v_c_condition} for $i=1,\ldots,N$, and any measurable $w:\mathbb{R}_{+}\to\mathbb{R}^{2N}$ satisfying \eqref{d_condition}, the solution of the system defined by \eqref{apply_system_oneorder} and \eqref{apply_ui_definition_N} satisfies $p(t)\in\mathcal{D}_s$ and 
\begin{align}
h(p(t))\ge\min\{\beta(h(p(0)),t),\chi(-\|w\|_{[0,t)}),c\}\label{apply_conclusion_w_N}
\end{align}
for $t\in[0,\infty)$, where
\begin{align}
h(p)&=\min\left\{\min_{i=1,\ldots,N}h_i(p),\min_{\substack{i,j=1,\ldots,N\\i\neq j}}h_{ij}(p)\right\}  
\end{align}
for $p\in\mathbb{R}^{2N}$, and
\begin{align}
&\mathcal{D}_s=
\nonumber\\
&\{p\in\mathbb{R}^{2N}:
|p_i-o_j|\ge d_o,\;
\forall\;i=1,\ldots,N,\; j=1,\ldots,N_o,
\nonumber\\
&\qquad |p_i-p_j|\ge d_a,\;
\forall\;i,j=1,\ldots,N,\; i\ne j\}.
\label{definition_D_s}
\end{align}
\end{theorem}

Theorem \ref{theorem_example_uncertain} is proved in Appendix \ref{proof_theorem_example_uncertain}.

\subsection{Numerical Examples}\label{subsection_numerical_Example}

Based on Theorem \ref{theorem_example_uncertain}, we conduct a numerical simulation for the case with 3 mobile agents and 10 obstacles. We still consider the mobile agents modeled with \eqref{apply_system_oneorder}, and the controller in the form of \eqref{apply_ui_definition_N}, where the command velocities are given by
\begin{subequations}
\begin{align}
v_1^c&=\frac{[8,8]^T-p_1}{|[8,8]^T-p_1|}\min\{0.1|[8,8]^T-p_1|,0.3\},\label{v_1_c}\\
v_2^c&=\frac{[-2,0]^T+p_1-p_2}{|[-2,0]^T+p_1-p_2|}\min\{|[-2,0]^T+p_1-p_2|,0.3\},\label{v_2_c}\\
v_3^c&=\frac{[0,-2]^T+p_1-p_3}{|[0,-2]^T+p_1-p_3|}\min\{|[0,-2]^T+p_1-p_3|,0.3\}.\label{v_3_c}
\end{align}
\end{subequations}
Each command velocity is defined as zero when its corresponding position error is zero. Here, agent $1$ is designated as the leader, with $v_1^c$ driving it to the destination point $[8,8]^T$. Agents $2$ and $3$ are followers, with $v_2^c$ and $v_3^c$ steering them to try to maintain formation offsets of $[-2,0]^T$ and $[0,-2]^T$ relative to the leader, respectively. Similar scenarios involving formation maintenance and collision avoidance have also been considered in \cite{Mestres_2024_RAL}.

\begin{figure*}[!t]
\centering
\begin{subfigure}[t]{0.185\textwidth}
    \centering
    \begin{overpic}[
        width=\linewidth,
        trim=1.5cm 0pt 1.5cm 0pt,
        clip
    ]{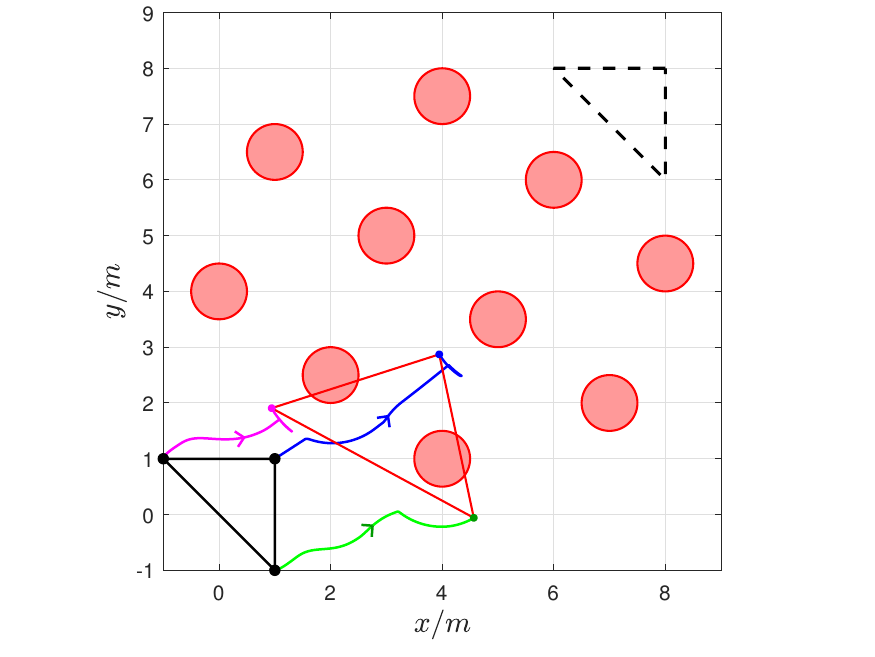}
    \end{overpic}
    \caption{$t=20\,\mathrm{s}$}
    \label{fig_trajectory_t70}
\end{subfigure}%
\begin{subfigure}[t]{0.185\textwidth}
    \centering
    \begin{overpic}[
        width=\linewidth,
        trim=1.5cm 0pt 1.5cm 0pt,
        clip
    ]{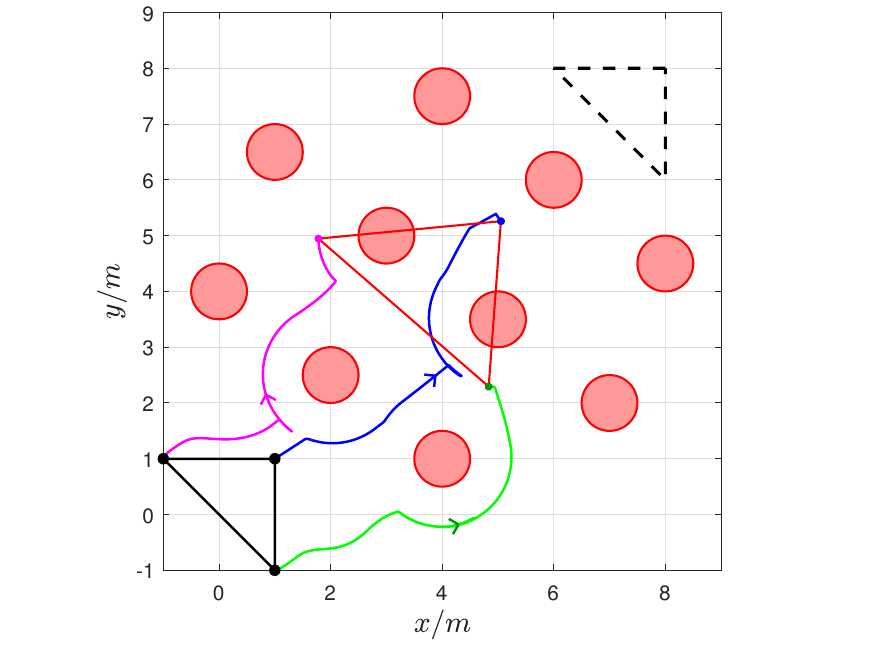}
    \end{overpic}
    \caption{$t=40\,\mathrm{s}$}
    \label{fig_trajectory_t140}
\end{subfigure}%
\begin{subfigure}[t]{0.185\textwidth}
    \centering
    \begin{overpic}[
        width=\linewidth,
        trim=1.5cm 0pt 1.5cm 0pt,
        clip
    ]{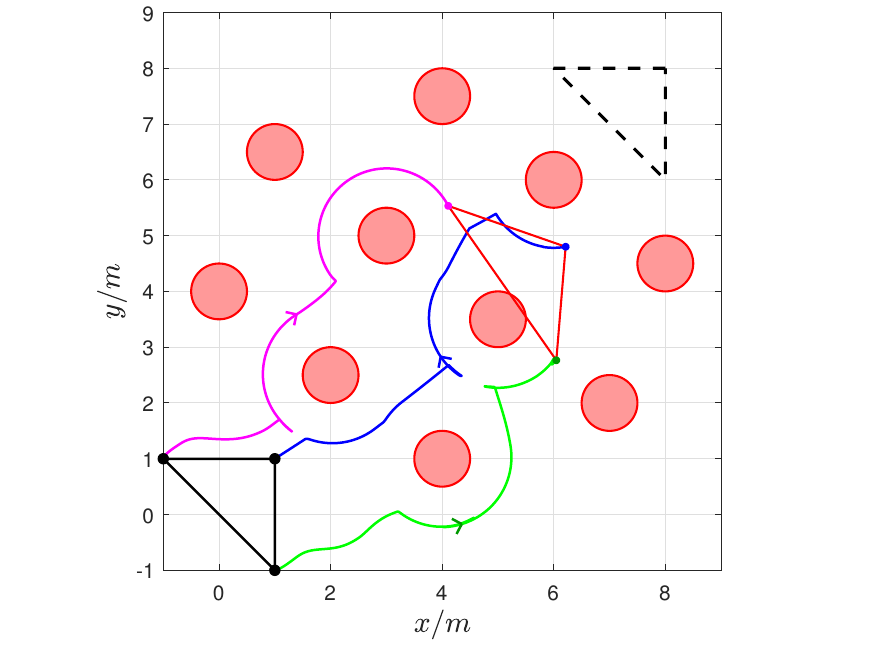}
    \end{overpic}
    \caption{$t=60\,\mathrm{s}$}
    \label{fig_trajectory_t210}
\end{subfigure}%
\begin{subfigure}[t]{0.185\textwidth}
    \centering
    \begin{overpic}[
        width=\linewidth,
        trim=1.5cm 0pt 1.5cm 0pt,
        clip
    ]{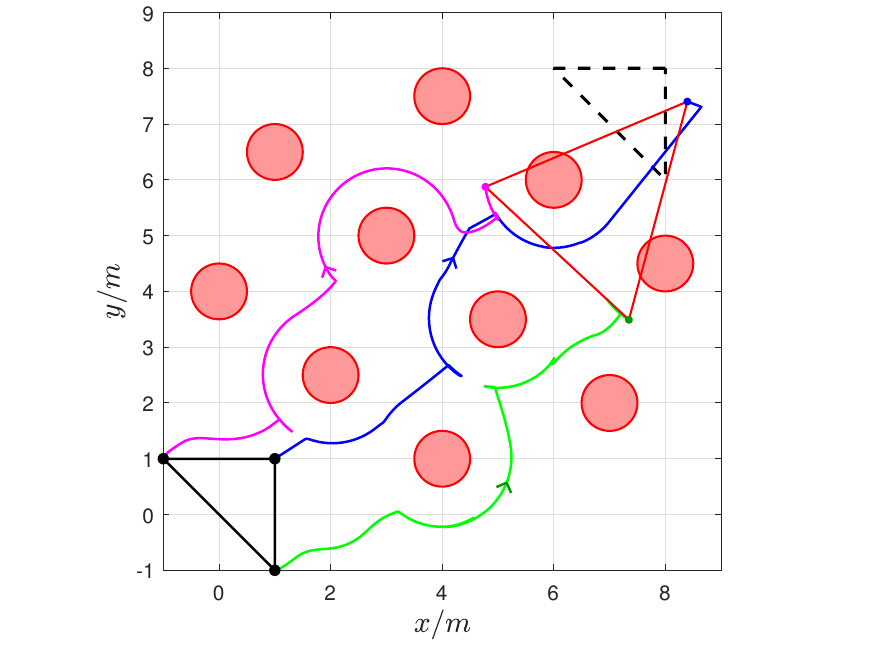}
    \end{overpic}
    \caption{$t=80\,\mathrm{s}$}
    \label{fig_trajectory_t280}
\end{subfigure}%
\begin{subfigure}[t]{0.185\textwidth}
    \centering
    \begin{overpic}[
        width=\linewidth,
        trim=1.5cm 0pt 1.5cm 0pt,
        clip
    ]{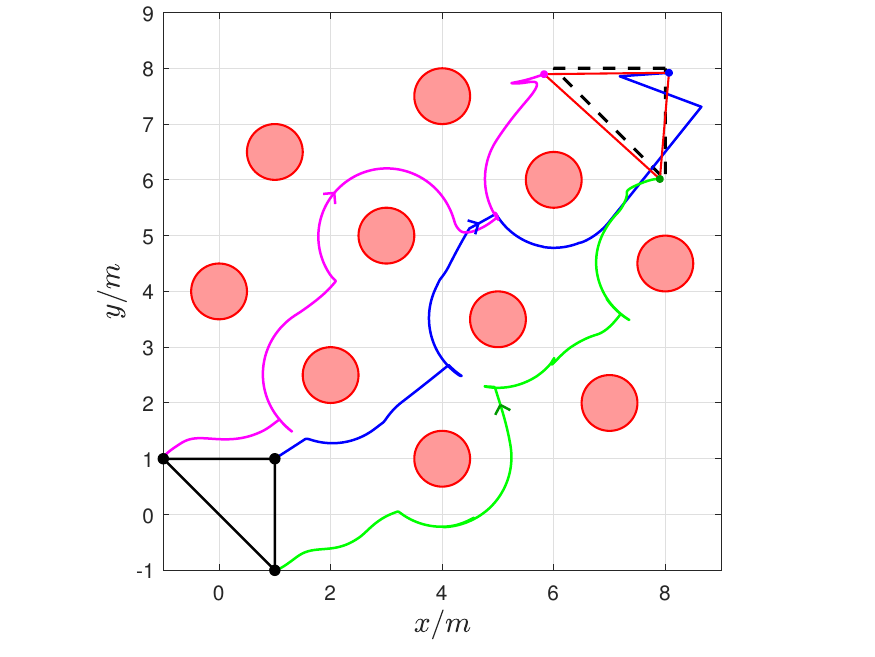}
    \end{overpic}
    \caption{$t=100\,\mathrm{s}$}
    \label{fig_trajectory_t350}
\end{subfigure}
\caption{The positions and trajectories of all mobile agents at
$t=20\,\mathrm{s}$, $t=40\,\mathrm{s}$, $t=60\,\mathrm{s}$,
$t=80\,\mathrm{s}$, and $t=100\,\mathrm{s}$ when $k_s=9$.}
\label{fig_simulation_result_N}
\end{figure*}

We set the desired minimum distance between the mobile agents and the obstacles to $d_o=0.5$, and the desired minimum distance between the mobile agents to $d_a=0.1$. As shown in Figure \ref{fig_simulation_result_N}, the ten obstacles are located at $[4,1]^T$, $[2,2.5]^T$, $[0,4]^T$, $[1,6.5]^T$, $[3,5]^T$, $[5,3.5]^T$, $[7,2]^T$, $[4,7.5]^T$, $[6,6]^T$, and $[8,4.5]^T$, respectively, and their minimum distance is $\min_{i\neq j}|o_i-o_j|=2.5$. According to the conditions in \eqref{parameter_choose_N}, the controller parameters are chosen as
\begin{subequations}\label{parameter_choose_N_numerical}
\begin{align}
\lambda&=0.2, &\quad k_s&=9, &\quad d_{mo}&=1.2465,\\
 d_{ma}&=0.7465,&\quad \mu&=0.99.
\end{align}
\end{subequations}

\begin{figure}[h!]
\centering
\begin{overpic}[width=0.9\linewidth]{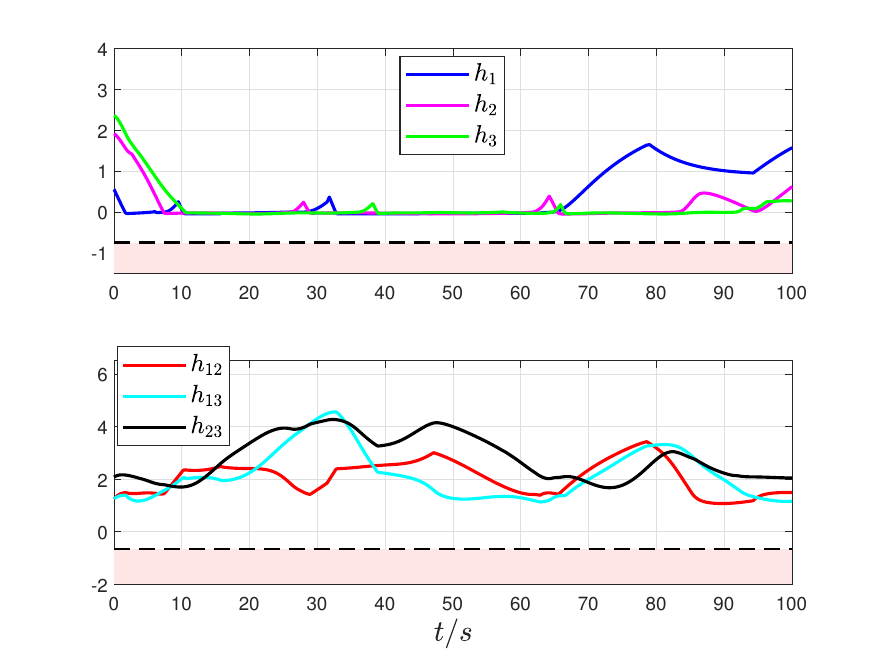}
\put(50,9.5){\scriptsize $\text{Unsafe Region}$}
\put(50,44.5){\scriptsize $\text{Unsafe Region}$}
\end{overpic}
\caption{Time histories of $h_1(p)$, $h_2(p)$, $h_3(p)$, $h_{12}(p)$, $h_{13}(p)$, $h_{23}(p)$ when $k_s=9$.}
\label{fig_simulation_result_N_safe_indicator_tra}
\end{figure}

Figures \ref{fig_simulation_result_N} and \ref{fig_simulation_result_N_safe_indicator_tra} present the simulation results for the initial positions
\begin{align}
p_1(0)=[1,1]^T,\quad p_2(0)=[-1,1]^T,\quad p_3(0)=[1,-1]^T
\end{align}
and the disturbances
\begin{subequations}
\begin{align}
w_1(t)&=0.12[\operatorname{sgn}(\sin(0.2t)),\,0]^T,\\
w_2(t)&=0.12[\sin(0.3t),\,\cos(0.3t)]^T,\\
w_3(t)&=0.12[\cos(0.4t),\,\sin(0.4t)]^T
\end{align}
\end{subequations}
for all $t\ge 0$, where $\operatorname{sgn}(s)$ equals $1$ for $s>0$, $0$ for $s=0$, and $-1$ for $s<0$. It can be observed that the mobile agents avoid collisions with both the obstacles and one another throughout the motion while attempting to maintain the desired formation, thereby validating the theoretical results.

\begin{figure}[h!]
\centering
\begin{overpic}[width=0.9\linewidth]{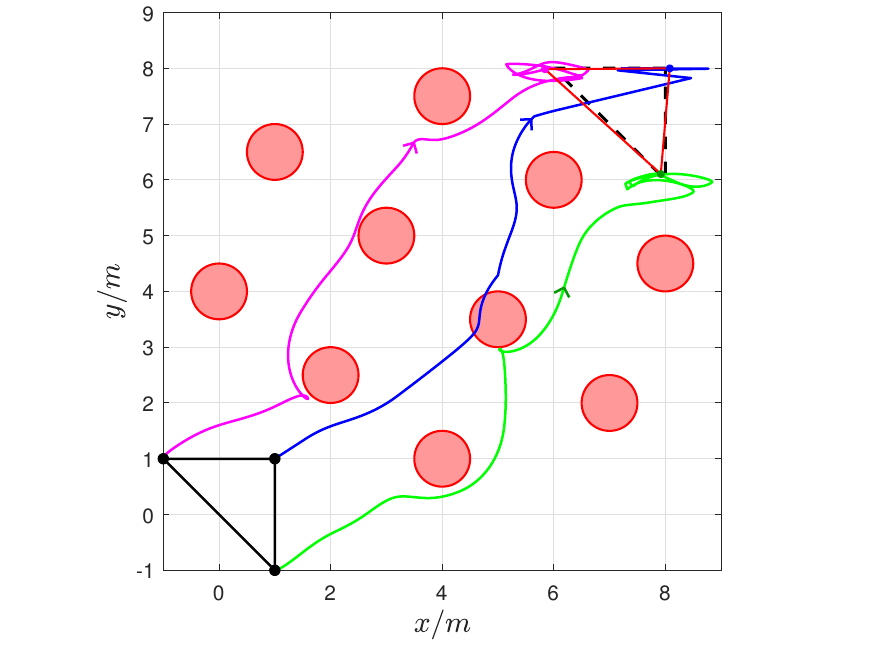}
\end{overpic}
\caption{The positions and trajectories of all mobile agents at $t=100\,\mathrm{s}$ when $k_s=0.3$.}
\label{fig_counter_simulation_result_N}
\end{figure}

\begin{figure}[h!]
\centering
\begin{overpic}[width=0.9\linewidth]{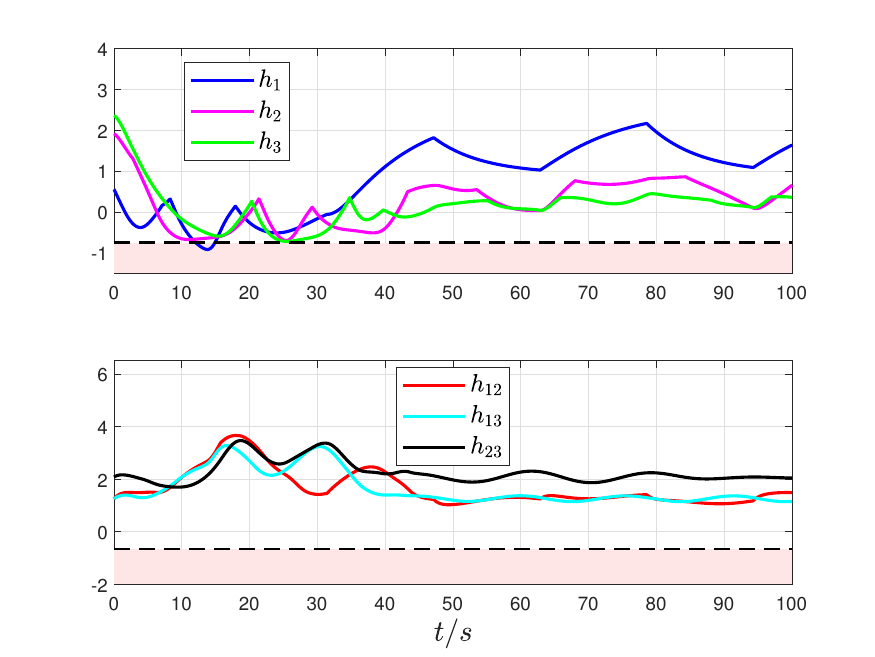}
\put(50,9.5){\scriptsize $\text{Unsafe Region}$}
\put(50,44.5){\scriptsize $\text{Unsafe Region}$}
\end{overpic}
\caption{Time histories of $h_1(p)$, $h_2(p)$, $h_3(p)$, $h_{12}(p)$, $h_{13}(p)$, $h_{23}(p)$ when $k_s=0.3$.}
\label{fig_counter_simulation_result_N_safe_indicator_tra}
\end{figure}

For comparison, we reduce $k_s$ to $0.3$ while keeping the remaining parameters unchanged from \eqref{parameter_choose_N_numerical}. This choice violates condition \eqref{k_s_choose_un} in Theorem \ref{theorem_example_uncertain}. The simulation results in Figures \ref{fig_counter_simulation_result_N} and \ref{fig_counter_simulation_result_N_safe_indicator_tra} show that the safety constraints between the agents and the obstacles are violated, illustrating the role of the proposed theoretical conditions in guiding the selection of controller parameters to ensure safety.

\section{Conclusion}\label{section_conclude}

In this paper, we have studied the safety verification problem for a general class of large-scale interconnected systems characterized by multiple safety indicator functions. Each safety indicator function satisfies a trajectory-based input-to-state safety (ISSf) estimate that provides a lower bound on its value along system trajectories in terms of the initial state, external inputs, and the other safety indicator functions, whose influence is quantified by continuous and nondecreasing gain functions. Our main contribution is a safety-oriented small-gain theorem and its corollaries that guarantee the desired safety properties of large-scale systems whenever their interconnection gain functions satisfy a small-gain condition on the negative real axis. This condition imposes restrictions on the interconnection gain functions only in the neighborhood of the safety boundary. To prove the result, we have built on key ideas and proof techniques from existing small-gain results and carefully analyzed and exploited the lower-bound structure of the ISSf estimates associated with the safety indicator functions. By combining these insights with an inductive comparison argument, we have established the ISSf property of the overall system throughout its maximal interval of existence. As an application, we have established robust safety for a class of distributed “half-spring” controllers, thereby providing a theoretical foundation for collision avoidance in multi-agent systems subject to obstacles and external disturbances.

It is our belief that the results of this paper provide a foundation for the systematic safety verification of broad classes of large-scale systems and open avenues for extensions to discrete-time, hybrid, and infinite-dimensional systems.

\appendices

\section{Convergence of a Class of Vector Systems with Self-Loops}\label{appendix_lemma_convergence}

Lemma \ref{Lemma_convergence_min} shows that, for inequality \eqref{self_loop_min} in the proof of Theorem \ref{theorem_smallgain_min}, provided the small-gain condition \eqref{circuitgain} holds, the self-loop term $\Gamma(\mathbf{H}_{[\mu t,t)})$ on the right-hand side can be removed by replacing the functions $\beta_i$ in $\mathbf{B}$ with a suitably smaller $\mathcal{K}^-\mathcal{L}$ function $\hat{\beta}$.

\begin{lemma}\label{Lemma_convergence_min}
Let $\mathbf{B}(\xi,t)=(\beta_1(\xi,t),\ldots,\beta_N(\xi,t))$, with $\beta_i\in\mathcal{K}^-\mathcal{L}$ for $i=1,\ldots,N$. Consider $\Gamma$ defined in \eqref{definition_Gamma_min} and \eqref{definition_Gamma_i_min} such that condition \eqref{circuitgain} is satisfied. Then, there exists $\hat{\beta}\in\mathcal{K}^-\mathcal{L}$ such that for any $\xi\le 0$, any $\mathbf{w} \in \mathbb{R}^{N}_{-}$, and any $\mathbf{z}(t) = (z_1(t),\ldots,z_N(t))$ with $z_i$ for
$i=1,\ldots,N$ being continuous and lower bounded on $[0,t_m)$, if
\begin{align}
\mathbf{z}(t)&\ge \operatorname{MIN}\{ \mathbf{B}(\xi,t),
\Gamma(\mathbf{z}_{[\mu t,t)}), \mathbf{w}\}\label{lemma_z_condition_min}
\end{align}
for all $t \in [0,t_m)$, then
\begin{align}
\lfloor\mathbf{z}(t)\rfloor \ge \min\{ \hat{\beta}(\xi,t), \lfloor Q(\mathbf{w})\rfloor \}\label{lemma_z_conclusion}
\end{align}
for all $t \in [0,t_m)$, where $\mu\in(0,1]$, $\mathbf{z}_{[\mu t,t)}=(\lfloor\!\lfloor z_1\rfloor\!\rfloor_{[\mu t,t)},\ldots,\lfloor\!\lfloor z_N\rfloor\!\rfloor_{[\mu t,t)})$, $t_m\in(0,\infty]$ and
\begin{align}
Q(\mathbf{s})=\operatorname{MIN}\{\mathbf{s},\Gamma_m(\mathbf{s}),[\Gamma_m]^2(\mathbf{s}),\ldots,[\Gamma_m]^{N-1}(\mathbf{s})\}\label{definition_Q_lemma}
\end{align}
for $\mathbf{s}\in\mathbb{R}^N$ with 
\begin{align}
\Gamma_m(\mathbf{s})=\operatorname{MIN}\{\Gamma(\mathbf{s}),\lambda\mathbf{s}\}\label{definition_Gamma_m_lemma}
\end{align}
and a constant $\lambda\in(0,1)$.
\end{lemma}

\begin{proof}
Let $t_0=0$ and $t_k\ge t_{k-1}$ be the first time instant such that 
\begin{align}
\mathbf{B}(\xi,t)\ge[\Gamma_{m}]^k(Q(\mathbf{B}(\xi,0)))\; \forall t\ge t_{k} \label{tl_definition}
\end{align}
for all integers $k\ge 1$. The existence of the sequence $\{t_k\}_{k\in\mathbb{Z}_+}$ is guaranteed by the fact that $\mathbf{B}$ is composed of $\mathcal{K}^-\mathcal{L}$ functions $\beta_i$. Define $\bar{t}_0=0$ and $\bar{t}_{k+1}=\max\{t_{k+1},\bar{t}_{k}/\mu\}$ for $k\in\mathbb{Z}_+$.

Denote $\bar{k}=\sup\{k\in\mathbb{Z}_+:\bar{t}_k< t_m\}$. We prove by induction that for any $k\le \bar{k}$, 
\begin{align}
\mathbf{z}(t)\ge \operatorname{MIN}\{[\Gamma_{m}]^k(Q(\mathbf{B}(\xi,0))),Q(\mathbf{w})\}\label{induction_convergence}
\end{align}	
holds for $t\in[\bar{t}_k,t_m)$. 

{Initial Step:}

We first show that \eqref{induction_convergence} holds for $k=0$. Denote
\begin{align}
\mathbf{z}_{[0,t)}=(\lfloor\!\lfloor z_1\rfloor\!\rfloor_{[0,t)},\ldots,\lfloor\!\lfloor z_N\rfloor\!\rfloor_{[0,t)}).
\end{align}
Under condition \eqref{lemma_z_condition_min}, it holds that
\begin{align}
\mathbf{z}_{[0,t)}&\ge \operatorname{MIN}\{\mathbf{B}(\xi,0),\Gamma(\mathbf{z}_{[0,t)}),\mathbf{w}\}.\label{z_0_t}
\end{align}
Substituting \eqref{definition_Gamma_m_lemma} into \eqref{z_0_t}, we have
\begin{align}
\mathbf{z}_{[0,t)}&\ge \operatorname{MIN}\{\mathbf{B}(\xi,0),
\Gamma_m(\mathbf{z}_{[0,t)}), \mathbf{w}\}.\label{z_Gamma_m}
\end{align}
By the definition of $\Gamma_m$ in \eqref{definition_Gamma_m_lemma}, the MIN-preserving property specified in \eqref{min_preserving1}--\eqref{min_preserving2} also holds for $\Gamma_m$. 

By substituting \eqref{z_Gamma_m} into itself and using the MIN-preserving property of $\Gamma_m$, we have
\begin{align}
\mathbf{z}_{[0,t)}&\ge\operatorname{MIN}\{\mathbf{B}(\xi,0),\mathbf{w},\nonumber\\
&~~~\Gamma_m(\operatorname{MIN}\{\mathbf{B}(\xi,0),\Gamma_m(\mathbf{z}_{[0,t)}),\mathbf{w}\})\}\nonumber\\
&\ge\operatorname{MIN}\{
\mathbf{B}(\xi,0), \Gamma_m(\mathbf{B}(\xi,0)), [\Gamma_m]^{2}(\mathbf{z}_{[0,t)}),\nonumber\\
&~~~\mathbf{w},\Gamma_m(\mathbf{w})\}.
\end{align}

Repeating the operation $(q^*-1)$ times results in 
\begin{align}
\mathbf{z}_{[0,t)}
&\ge \operatorname{MIN}\{\mathbf{B}(\xi,0), \Gamma_m(\mathbf{B}(\xi,0)),\ldots,[\Gamma_m]^{q^*-1}(\mathbf{B}(\xi,0)),\nonumber\\
&~~~[\Gamma_m]^{q^*}(\mathbf{z}_{[0,t)}),\mathbf{w},\Gamma_m(\mathbf{w}),\ldots,[\Gamma_m]^{q^*-1}(\mathbf{w})\}.\label{z_m_qstar}
\end{align}

By repeating the reasoning from \eqref{definition_Qbar} to \eqref{H_m_lower_bound}, we have, for each $k\in\mathbb{Z}_+$,
\begin{align}
[\Gamma_{m}]^{k}(\mathbf{s}) \ge Q(\mathbf{s})\label{Gamma_m_k_lowerbound}
\end{align}
for all $\mathbf{s}\in\mathbb{R}^N$, and
\begin{align}
\mathbf{z}_{[0,t)}\ge  \operatorname{MIN}\{
Q(\mathbf{B}(\xi,0)), Q(\mathbf{w})\}\label{lowerboundness_z}
\end{align}
for all $t\in[0,t_m)$, which implies \eqref{induction_convergence} for $k=0$.

{Inductive Step:}

Then, we suppose that property \eqref{induction_convergence} holds for $k=q<\bar{k}$, and we show that \eqref{induction_convergence} holds for $k=q+1$. Substituting \eqref{induction_convergence} with $k=q$ and \eqref{tl_definition} with $k=q+1$ into \eqref{lemma_z_condition_min}, we have
\begin{align}
\mathbf{z}(t)&\ge \operatorname{MIN}\{[\Gamma_{m}]^{q+1}(Q(\mathbf{B}(\xi,0))),\nonumber\\
&~~~\Gamma_{m}(Q(\mathbf{w})),\mathbf{w}\}\label{z_lowerbound_Gamma_m}
\end{align}
for $t\in [\bar{t}_{q+1},t_m)$. With the definition of $Q$ in \eqref{definition_Q_lemma}, by using property \eqref{Gamma_m_k_lowerbound}, for each integer $k\ge 0$,
\begin{align}
[\Gamma_{m}]^{k}(Q(\mathbf{s})) \ge Q(\mathbf{s})\label{Gamma_m_k_lowerbound_Q}
\end{align}
for $\mathbf{s}\in\mathbb{R}^N$. Substituting \eqref{Gamma_m_k_lowerbound_Q} into \eqref{z_lowerbound_Gamma_m}, we have
\begin{align}
\mathbf{z}(t)\ge\operatorname{MIN}\{[\Gamma_{m}]^{q+1}(Q(\mathbf{B}(\xi,0))),Q(\mathbf{w})\}
\end{align}
for $t\in [\bar{t}_{q+1},t_m)$, which implies property \eqref{induction_convergence} by induction. The basic idea of the claim is shown in Figure \ref{fig_proof}.

\begin{figure}[h!]
\centering
\includegraphics[width=0.9\linewidth]{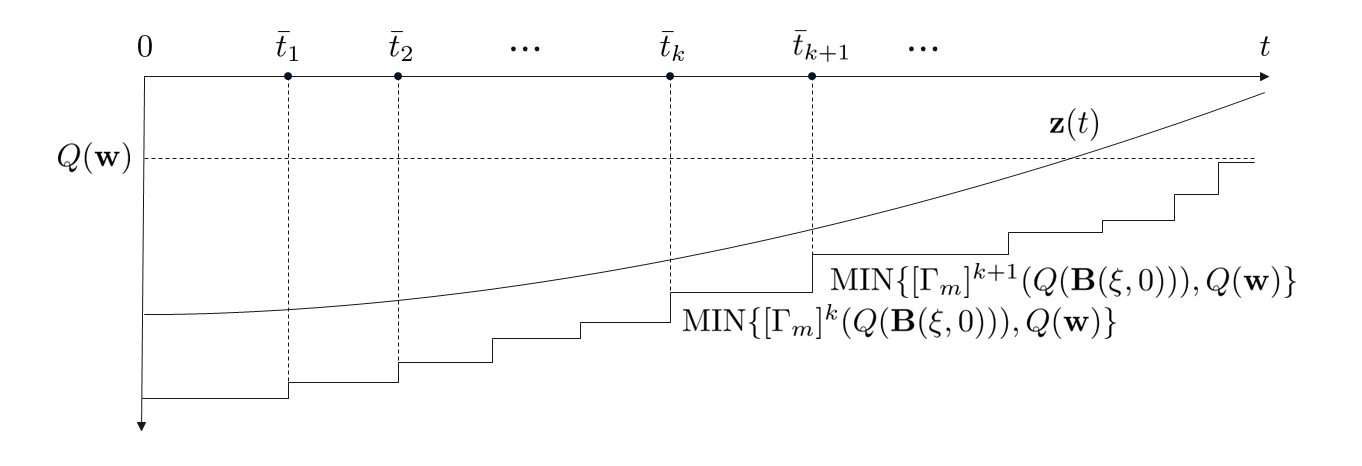}
\caption{Illustration of the time sequence $\bar{t}_k$ and the corresponding lower bound $\operatorname{MIN}\{[\Gamma_m]^{k}(Q(\mathbf{B}(\xi,0))),Q(\mathbf{w})\}$ in the induction.}
\label{fig_proof}
\end{figure}

For safety-critical systems, forward completeness is not
ensured and the signal $\mathbf{z}$ may not be defined on $[0,\infty)$. Based
on property \eqref{induction_convergence}, to construct a lower bound for $\mathbf{z}(t)$ in terms of a $\mathcal{K}^-\mathcal{L}$ function, we define an auxiliary signal vector $\mathbf{v}$ which is defined on $[0,\infty)$ for comparison.

Let $\mathbf{v}:\mathbb{R}_{+}\to\mathbb{R}^N$ be a continuous signal such that $\mathbf{v}(t)\le\mathbf{z}(t)$ for all $t\in[0,t_m)$ and
\begin{align}
\mathbf{v}(t)\ge \operatorname{MIN}\{[\Gamma_{m}]^k(Q(\mathbf{B}(\xi,0))),Q(\mathbf{w})\}\label{induction_convergence_v}
\end{align}	
for $t\in[\bar{t}_k,\infty)$. Property \eqref{induction_convergence} guarantees the existence of such $\mathbf{v}$. 

Following the same reasoning as in
\eqref{definition_Qbar}--\eqref{Gammabar_k_convergence}, we conclude that, for each $\epsilon<0$, there exists an $N_\epsilon\in\mathbb{Z}_+$ such that $[\Gamma_{m}]^k(Q(\mathbf{B}(\xi,0)))\ge \epsilon\mathbf{1}_N$ for all $k\ge N_\epsilon$, and thus, with \eqref{induction_convergence_v}, for any $\mathbf{w}\in\mathbb R_-^N$,
\begin{align}
\lfloor \mathbf{v}(t) \rfloor\ge \min\{\epsilon,\lfloor Q(\mathbf{w})\rfloor\}
\end{align}
for $t\in[\bar{t}_k,\infty)$ with $k\ge N_\epsilon$. Moreover, for each $\epsilon<0$, there exists a $\delta>0$ such that
\begin{align}
\lfloor\mathbf{v}(t)\rfloor\ge \epsilon,\quad \forall t\ge 0
\end{align}
for any $\mathbf{w}$ satisfying $\lfloor \mathbf{w}\rfloor\ge -\delta$ and any $\xi$ satisfying $\xi\ge -\delta$. Then, following an argument similar to that used in
\cite[Lemma 2.7]{SONTAG_1995_SCL}, we obtain a function
$\hat{\beta}\in\mathcal{K}^-\mathcal{L}$ such that
\begin{align}
\lfloor\mathbf{v}(t)\rfloor \ge \min\{ \hat{\beta}(\xi,t), \lfloor Q(\mathbf{w})\rfloor \}
\end{align}
for $t\ge 0$, which implies \eqref{lemma_z_conclusion}.
\end{proof}

\section{Proof of Proposition \ref{proposition_controller_equivalence}}\label{proof_proposition_controller_equivalence}
With the definition of $v_{ij}^s$ in \eqref{definition_v_s_ij}, it is straightforward to verify that
\begin{align}
\sum_{\substack{j=1,\ldots,N\\ j\neq i}}v^s_{ij}(p)=\sum_{\substack{j=1,\ldots,N\\ j\neq i}}\frac{p_i-p_j}{|p_i-p_j|}\min\{|p_i-p_j|-d_{ma},0\}\label{h_ij_equal}
\end{align}
for all $p\in\mathcal{D}$. Next, we show that the first term of the controller \eqref{apply_ui_definition_N} equals the first term of the controller \eqref{apply_ui_definition_N_equivalent}. Under condition \eqref{d_mo_condition_1}, when $h_i(p)< 0$, there exists only one $j$ such that $|p_i-o_j|-d_{mo}<0$. Thus, by the definition of $h_i$ in \eqref{definition_h_motivalexample}, when $h_i(p)<0$ and $p\in\mathcal{D}$, we have $h_i(p)=|p_i-o_{S_i(p)}|-d_{mo}$, which implies that
\begin{align}
\frac{\partial h_i(p)}{\partial p_i}=\frac{\partial |p_i-o_{S_i(p)}|}{\partial p_i}=\frac{p_i-o_{S_i(p)}}{|p_i-o_{S_i(p)}|}\label{gradient_h_i_example}
\end{align}
and
\begin{align}
\sum_{j=1}^{N_o}\frac{p_i-o_j}{|p_i-o_j|}\min\{|p_i-o_j|-d_{mo},0\}=\frac{\partial h_i(p)}{\partial p_i}h_i(p),\label{h_i_equal}
\end{align}
where
\begin{align}
S_i(p)=\min\{j:|p_i-o_j|-d_{mo}=h_i(p)\}.\label{definition_I_i_example}
\end{align}
Combining \eqref{h_ij_equal} and \eqref{h_i_equal}, we conclude that the two controllers \eqref{apply_ui_definition_N} and \eqref{apply_ui_definition_N_equivalent} are equivalent.

\section{Proof of Proposition \ref{prop_robust_safety_controller}}\label{proof_prop_robust_safety_controller}

We first establish the ISSf estimate for $h_i(p(t))$ and then that for $h_{ij}(p(t))$.

\subsection{Proof of the ISSf Estimate for $h_i$}

For each $i=1,\ldots,N$, when $h_i(p)< 0$, along the trajectories of \eqref{apply_system_oneorder} and \eqref{apply_ui_definition_N}, we have
\begin{align}
&(\frac{\partial h_i(p)}{\partial p})^T\dot{p}\nonumber\\
&\ge -k_s h_{i}(p)+k_s \lambda\sum_{j\neq i}\min\{|p_i-p_j|-d_{ma},0\}\nonumber\\
&~~~-\bar{v}_c-\|w_i\|_{[0,\infty)}\nonumber\\
&= -k_s h_i(p)-\bar{v}_c-\|w_i\|_{[0,\infty)}\nonumber\\
&~~~+k_s \lambda\sum_{j\neq i}\min\{|p_i-o_{S_i(p)}-p_j+o_{S_i(p)}|-d_{ma},0\}\nonumber\\
&\ge -k_s h_i(p)-\bar{v}_c-\|w_i\|_{[0,\infty)}\nonumber\\
&~~~+k_s \lambda\sum_{j\neq i}\min\{h_{j}(p)-h_{i}(p)-d_{ma},0\}\nonumber\\
&\ge -\epsilon  h_i(p)-(k_s-\epsilon) h_i(p)\nonumber\\
&~~~+(N+1)\min_{j\neq i}\{k_s \lambda(2\min\{h_{j}(p),-d_{ma}\}-h_{i}(p)),\nonumber\\
&~~~-\bar{v}_c,-\|w_i\|_{[0,\infty)}\}\label{h_i_d_cal}
\end{align} 
for all $p\in\mathcal{D}$, where $\epsilon\in(0,k_s)$ is a constant and $S_i(p)$ is defined in \eqref{definition_I_i_example}. Therefore, we have
\begin{align}
&h_i(p)\le \min_{j\neq i}\{\hat{\gamma}_{ij}(h_{j}(p)),\chi_o(-\|w_i\|_{[0,\infty)}),c_o\}
\nonumber\\
\Rightarrow&(\frac{\partial h_i(p)}{\partial p})^T\dot{p}\ge -\epsilon h_i(p)\label{ISSf_h_i_apply_N}
\end{align}
where
\begin{subequations}
\begin{align}
\hat{\gamma}_{ij}(s)&=\min\left\{\frac{2(N+1)k_s \lambda}{k_s-\epsilon+(N+1)k_s \lambda}s,0\right\},\label{definition_gamma_ij_example}\\
\chi_o(s)&=\min\left\{\frac{N+1}{k_s-\epsilon}s,0\right\}\label{definition_chi_o}\\
c_o&=-\max\left\{\frac{2(N+1)k_s \lambda d_{ma}}{k_s-\epsilon+(N+1)k_s \lambda},\frac{(N+1)\bar{v}_c}{k_s-\epsilon}\right\},\label{definition_c_o}
\end{align}
\end{subequations}
for $s\in\mathbb{R}$. By Theorem \ref{ISSf-BFtheorem} and causality,
there exists $\beta_o\in\mathcal{K}^-\mathcal{L}$ such that
\eqref{interconnection_N} holds and $\beta_o(s,0)=s$
for all $s\le 0$.

\subsection{Proof of the ISSf Estimate for $h_{ij}$}

For each $i,j=1,\ldots,N$, $j\neq i$, along the trajectories of \eqref{apply_system_oneorder} and \eqref{apply_ui_definition_N}, direct calculation yields the following inequality:
\begin{align}
&(\frac{\partial h_{ij}(p)}{\partial p})^T\dot{p}\nonumber\\
&\ge -2k_s \lambda \min\{h_{ij}(p),0\}-\|w_i\|_{[0,\infty)}-\|w_j\|_{[0,\infty)}-2\bar{v}_c\nonumber\\
&~~~-k_s \min\{h_{i}(p),0\}\cos\theta_1-k_s \min\{h_{j}(p),0\}\cos\theta_2\nonumber\\
&~~~-k_s \lambda\sum_{k\neq i,j}(\cos\theta_{ijk}\min\{h_{ik}(p),0\}\nonumber\\
&~~~+ \cos\theta_{jik}\min\{h_{jk}(p),0\})\label{h_12_derivative_theta}
\end{align}
for all $p\in\mathcal{D}$, where
\begin{subequations}
\begin{align}
\theta_1&=\arccos\left(\frac{(p_i-p_j)^T}{|p_i-p_j|}\frac{p_i-o_{S_i(p)}}{|p_i-o_{S_i(p)}|}\right),\label{definition_theta_1}\\
\theta_2&=\arccos\left(\frac{(p_j-p_i)^T}{|p_j-p_i|}\frac{p_j-o_{S_j(p)}}{|p_j-o_{S_j(p)}|}\right),\label{definition_theta_2}\\
\theta_{ijk}&=\arccos\left(\frac{(p_i-p_j)^T}{|p_i-p_j|}\frac{p_i-p_k}{|p_i-p_k|}\right), \nonumber\\
&~~~i,j,k\in\{1,\ldots,N\},\;
i\neq j,\;i\neq k,\;j\neq k\label{definition_theta_ijk}
\end{align}
\end{subequations}
with $S_i$ and $S_j$ defined in \eqref{definition_I_i_example}.

For some terms on the right-hand side of inequality \eqref{h_12_derivative_theta}, we have the following claim.

\textbf{Claim:} For any $p\in\mathcal{D}$, it holds that
\begin{align}
&-k_s\min\{h_{i}(p),0\}\cos\theta_1 -k_s \min\{h_{j}(p),0\}\cos\theta_2-\nonumber\\
&k_s \lambda\sum_{k\neq i,j}\left(\cos\theta_{ijk}\min\{h_{ik}(p),0\}+ \cos\theta_{jik}\min\{h_{jk}(p),0\}\right)\nonumber\\
&\ge -k_s|p_i-p_j|(\lambda(N-2)+1).\label{claim_conclusion}
\end{align}

The proof of the claim can be seen in \eqref{theta_12_term}--\eqref{h_ik_sumh_jk_cos}.

Whenever $h_{ij}(p)\le -d_{ma}+d_a$, it holds that
\begin{align}
-|p_i-p_j|\ge -d_a\ge \frac{d_a}{d_{ma}-d_a}h_{ij}(p). \label{p_ij_lower}
\end{align}

Substituting \eqref{claim_conclusion} and \eqref{p_ij_lower} into \eqref{h_12_derivative_theta}, we have
\begin{align}
h_{ij}(p)&\le -d_{ma}+d_a\Rightarrow\nonumber\\
(\frac{\partial h_{ij}(p)}{\partial p})^T\dot{p}&\ge -2k_s \lambda h_{ij}(p)-\|w_i\|_{[0,\infty)}-\|w_j\|_{[0,\infty)}\nonumber\\
&~~~-k_s(\lambda(N-2)+1)|p_i-p_j|-2\bar{v}_c\nonumber\\
&\ge -2k_s \lambda h_{ij}(p)-\|w_i\|_{[0,\infty)}-\|w_j\|_{[0,\infty)}\nonumber\\
&~~~+k_s(1+\lambda(N-2)) \frac{d_a}{d_{ma}-d_a}h_{ij}(p)-2\bar{v}_c\nonumber\\
&\ge-\epsilon_a h_{ij}(p)-(k_a-\epsilon_a)h_{ij}(p)\nonumber\\
&~~~-4\max\{\|w_i\|_{[0,\infty)},\|w_j\|_{[0,\infty)},\bar{v}_c\}
\end{align}
for $p\in\mathcal{D}$, where the constant $k_a$ is defined by
\begin{align}
k_a=2k_s\lambda-\frac{k_s(1+\lambda(N-2))d_a}{d_{ma}-d_a}\label{definition_k_a}
\end{align}
and $\epsilon_a\in(0,k_a)$. Then, 
\begin{align}
&h_{ij}(p)\le \min\left\{-\frac{4}{k_a-\epsilon_a}\|w_i\|_{[0,\infty)},-\frac{4}{k_a-\epsilon_a}\|w_j\|_{[0,\infty)},\right.\nonumber\\
&\left.-\frac{4\bar{v}_c}{k_a-\epsilon_a},-d_{ma}+d_a\right\}\Rightarrow
(\frac{\partial h_{ij}(p)}{\partial p})^T\dot{p}\ge-\epsilon_a h_{ij}(p).\label{h_12_lower_s}
\end{align}

Under condition \eqref{d_ma_condition}, $k_a>0$, and $\epsilon_a$ can be chosen to satisfy $\epsilon_a\in(0,k_a)$. Therefore, based on property \eqref{h_12_lower_s}, by Theorem \ref{ISSf-BFtheorem} and causality, there exists $\beta_a\in\mathcal{K}^-\mathcal{L}$ such that property \eqref{h_ij_robust} holds and $\beta_a(s,0)=s$ for $s\le 0$, where
\begin{align}
\chi_a(s)&=\min\left\{\frac{4}{k_a-\epsilon_a}s,0\right\},\label{definition_chi_a}\\
c_a&=\min\left\{-\frac{4\bar{v}_c}{k_a-\epsilon_a},-d_{ma}+d_a\right\}\label{definition_c_a}
\end{align}
for $s\in\mathbb{R}$.

\textbf{Proof of the claim:} We first show that
\begin{align}
&-k_s\min\{h_{i}(p),0\}\cos\theta_1 -k_s \min\{h_{j}(p),0\}\cos\theta_2\nonumber\\
&\ge -k_s|p_i-p_j|. \label{theta_12_term}
\end{align}
If $\min\{h_i(p),h_j(p)\}\ge 0$, inequality \eqref{theta_12_term} holds. Hence, it suffices to consider the case where $\min\{h_i(p),h_j(p)\}<0$. Without loss of generality, let $h_i(p)<0$. We then distinguish between $h_j(p)\ge 0$ and $h_j(p)<0$.

When $h_j(p)\ge 0$, it holds that for each $k=1,\ldots,N_o$, 
\begin{align}
|p_j-o_k|\ge d_{mo}.\label{p_j_o_k}
\end{align}
With property \eqref{p_j_o_k}, we have
\begin{align}
|p_i-o_{S_i(p)}|\ge |p_j-o_{S_i(p)}|-|p_i-p_j|\ge d_{mo}-|p_i-p_j|. \label{p_j_o_k_condition}
\end{align}
By the definition of $h_i$ in \eqref{definition_h_motivalexample} and property \eqref{p_j_o_k_condition}, we have
\begin{align}
&-k_s\min\{h_{i}(p),0\}\cos\theta_1 -k_s \min\{h_{j}(p),0\}\cos\theta_2\nonumber\\
&\ge k_s\min\{|p_i-o_{S_i(p)}|-d_{mo},0\}\ge -k_s|p_i-p_j|.
\end{align}

When $h_j(p)< 0$ and $S_i(p)=S_j(p)$, by the definition of $\theta_1$ in \eqref{definition_theta_1} and $\theta_2$ in \eqref{definition_theta_2}, it holds that $\theta_1+\theta_2\le \pi$. Then, we have
\begin{align}
&-k_s \min\{h_{i}(p),0\}\cos\theta_1 -k_s \min\{h_{j}(p),0\}\cos\theta_2\nonumber\\
&\ge -k_s h_{i}(p)\cos\theta_1 -k_s h_{j}(p)\cos(\pi-\theta_1)\nonumber\\
&=-k_s(h_{i}(p)-h_j(p))\cos\theta_1.\label{term_less}
\end{align}
Substituting the definition of $h_i$ in \eqref{definition_h_motivalexample} into \eqref{term_less}, we have
\begin{align}
&-k_s \min\{h_{i}(p),0\}\cos\theta_1 -k_s \min\{h_{j}(p),0\}\cos\theta_2\nonumber\\
&\ge -k_s(|p_i-o_{S_i(p)}|-|p_j-o_{S_j(p)}|)\cos\theta_1\nonumber\\
&=-k_s(|p_i-o_{S_i(p)}|-|p_j-o_{S_i(p)}|)\cos\theta_1\ge -k_s|p_i-p_j|.
\end{align}
When $h_j(p)< 0$ and $S_i(p)\neq S_j(p)$, by the definition of $h_i$ in \eqref{definition_h_motivalexample}, it holds that
\begin{align}
&-k_s \min\{h_{i}(p),0\}\cos\theta_1 -k_s \min\{h_{j}(p),0\}\cos\theta_2\nonumber\\
&\ge k_s h_{i}(p) +k_s h_{j}(p)\nonumber\\
&=k_s(|p_i-o_{S_i(p)}|+|p_j-o_{S_j(p)}|-2d_{mo}).\label{term_less_2}
\end{align}
Substituting 
\begin{align}
|o_{S_i(p)}-p_i|+|p_i-p_j|+|p_j-o_{S_j(p)}|\ge |o_{S_i(p)}-o_{S_j(p)}|
\end{align}
and \eqref{d_mo_condition_1} into \eqref{term_less_2}, we have
\begin{align}
&-k_s \min\{h_{i}(p),0\}\cos\theta_1 -k_s \min\{h_{j}(p),0\}\cos\theta_2\nonumber\\
&\ge k_s(|o_{S_i(p)}-o_{S_j(p)}|-|p_i-p_j|-2d_{mo})\nonumber\\
&\ge -k_s|p_i-p_j|.\label{term_less_3}
\end{align}

By an argument similar to that used in the derivation from \eqref{theta_12_term} to \eqref{term_less_3}, it holds that for each $k=1,\ldots,N$ and $k\neq i,j$,
\begin{align}
&-k_s \lambda \cos\theta_{ijk}\min\{h_{ik}(p),0\}-k_s \lambda \cos\theta_{jik}\min\{h_{jk}(p),0\}\nonumber\\
&\ge  -k_s \lambda|p_i-p_j|.\label{h_ik_sumh_jk_cos}
\end{align}
Combining properties \eqref{theta_12_term} and \eqref{h_ik_sumh_jk_cos}, we have \eqref{claim_conclusion}.

\section{Proof of Theorem \ref{theorem_example_uncertain}}\label{proof_theorem_example_uncertain}

We first establish that the system defined by \eqref{apply_system_oneorder} and \eqref{apply_ui_definition_N} is ISSf with respect to $h_i$, and then prove that its safety constraints are satisfied.

\subsection{Proof of ISSf of the System Defined by \eqref{apply_system_oneorder} and \eqref{apply_ui_definition_N}}

Under conditions \eqref{condition_distance_un} and
\eqref{c_tilde_N}, it holds that
\[
0<\tilde{\lambda}<\frac{1}{N+1}.
\]
With \eqref{d_ma_choose_un}, we have
\begin{align}
d_{ma}
&=\frac{N+1}{1-(N+1)\tilde{\lambda}}d_a\nonumber\\
&=(N+1)d_a+\frac{\mu}{2}
(\min_{i\neq j}|o_i-o_j|-2d_o-2(N+1)d_a).
\end{align}
Since $\mu\in(0,1)$, condition \eqref{d_mo_choose_un}
implies that
\begin{align}
2d_o<2d_{mo}
=2d_o+2d_{ma}
<\min_{i\neq j}|o_i-o_j|,
\end{align}
which verifies \eqref{d_mo_condition_1}.
Moreover, under condition \eqref{c_choose_un},
\begin{align}
2\lambda d_{ma}
&=\frac{2(N+1)\lambda}{1-(N+1)\tilde{\lambda}}d_a
>2d_a
>(1+N\lambda)d_a.
\end{align}
Thus, condition \eqref{d_ma_condition} also holds.

Under conditions \eqref{d_mo_condition_1} and \eqref{d_ma_condition}, using Proposition \ref{prop_robust_safety_controller}, we have properties \eqref{interconnection_N} and \eqref{h_ij_robust}, which imply that Assumptions \ref{assum_ISSf} and \ref{assum_UO} hold.

We first establish safety between the agents and the obstacles, and then establish safety between the agents.

Under condition \eqref{c_choose_un}, with the definition of $\hat{\gamma}_{ij}$ in \eqref{definition_gamma_ij_example}, we can choose
\begin{align}
\epsilon\in(0,k_s-(N+1)k_s\lambda)\label{epsilon_choosen}
\end{align}
such that
\begin{align}
\hat{\gamma}_{ij}(s)=\frac{2(N+1)k_s \lambda}{k_s-\epsilon+(N+1)k_s \lambda}s>s\label{gamma_hat_smallgain}
\end{align}
for all $s< 0$. Therefore, the small-gain condition \eqref{circuitgain} is satisfied: along the simple cycle $(i_1,i_2,\ldots,i_r,i_1)$,
\begin{align}
\hat{\gamma}_{i_1 i_2}\circ\hat{\gamma}_{i_2 i_3}\circ\cdots\circ \hat{\gamma}_{i_{r-1} i_r}\circ \hat{\gamma}_{i_r i_1}(s)&>s,\quad\forall s<0,\label{circuitgain3_example_un}
\end{align}
for $r =2,\ldots,N$, $i_j \in \{1,\ldots,N\}$, $i_j\neq i_{j'}$ if $j\neq j'$.

Since Assumptions \ref{assum_ISSf} and \ref{assum_UO} hold, with small-gain condition \eqref{circuitgain} satisfied, by Theorem \ref{theorem_smallgain_min}, there exist $\beta\in\mathcal{K}^-\mathcal{L}$, $\chi\in\mathcal{K}^-$ and a constant $c\le 0$ such that
\begin{align}
\min_{i=1,\ldots,N}h_i(p(t))&\ge\min\{\beta(\min_{i=1,\ldots,N}h_i(p(0)),t),\nonumber\\
&~~~\chi(-\|w\|_{[0,t)}),c\}\label{h_i_convergence}
\end{align}
for all $t\in[0,t_m)$, and 
\begin{align}
h_i(p(t))&\ge \min\{\beta_o(h_i(p(0)),0),\min_{j\neq i} \hat{\gamma}_{ij}(\beta_o(h_j(p(0)),0)),\nonumber\\
&~~~c_o,\chi_o(-\|w_i\|_{[0,t)}),\min_{j\neq i}\hat{\gamma}_{ij}\circ\chi_o(-\|w_j\|_{[0,t)})\}\label{h_i_bound_example}
\end{align}
for all $t\in[0,t_m)$. 

Combining \eqref{h_i_convergence} and \eqref{h_ij_robust}, we conclude that property \eqref{apply_conclusion_w_N} holds for all $t\in[0,t_m)$.

\subsection{Proof of Safety Constraint Satisfaction of the System Defined by \eqref{apply_system_oneorder} and \eqref{apply_ui_definition_N}}

With $\beta_o(s,0)=s$ for all $s\le 0$ and $p(0)\in\mathcal{D}_s$, substituting \eqref{d_condition}, \eqref{gamma_hat_smallgain}, and the definition of $\mathcal{D}_s$ in \eqref{definition_D_s} into \eqref{h_i_bound_example}, we have
\begin{align}
h_i(p(t))&\ge \min\{d_{o}-d_{mo},c_o,\chi_o(-\bar{w})\}.\label{h_i_bound_2_example}
\end{align}
With the definitions of $c_o$ in \eqref{definition_c_o} and $\chi_o$ in \eqref{definition_chi_o}, under conditions \eqref{c_choose_un}, \eqref{k_s_choose_un}, and \eqref{d_mo_choose_un}, for any $\epsilon$ satisfying \eqref{epsilon_choosen} it holds that 
\begin{align}
\min\{c_o,\chi_o(-\bar{w})\}\ge d_{o}-d_{mo}.\label{c_o_bound}
\end{align}
Substituting \eqref{c_o_bound} into \eqref{h_i_bound_2_example}, we have
\begin{align}
h_i(p(t))&\ge d_{o}-d_{mo}\label{h_i_safe}
\end{align}
for all $t\in[0,t_m)$.

According to property \eqref{h_ij_robust}, with
$\beta_a(s,0)=\min\{s,0\}$ and the monotonicity of
$\beta_a(s,\cdot)$ for $s<0$, it holds that
\begin{align}
h_{ij}(p(t))
&\ge\min\{
h_{ij}(p(0)),\chi_a(-\|w_i\|_{[0,t)}),\nonumber\\
&~~~\chi_a(-\|w_j\|_{[0,t)}),c_a\}
\label{h_ij_bound_example}
\end{align}
for all $t\in[0,t_m)$. By \eqref{definition_k_a} and
\eqref{k_s_choose_un}, it holds that
\begin{align}
k_a(d_{ma}-d_a)
&=k_s(2\lambda d_{ma}-(1+N\lambda)d_a)
\nonumber\\
&>4\max\{\bar v_c,\bar w\}.
\end{align}
Hence, we can choose
\begin{align}
\epsilon_a\in\left(
0,k_a-\frac{4\max\{\bar v_c,\bar w\}}{d_{ma}-d_a}
\right),
\label{epsilon_a_choosen}
\end{align}
which implies that
\begin{align}
\frac{4\max\{\bar v_c,\bar w\}}{k_a-\epsilon_a}
<d_{ma}-d_a.
\end{align}
With the definitions of $\chi_a$ and $c_a$ in
\eqref{definition_chi_a} and \eqref{definition_c_a},
we obtain
\begin{align}
&\min\{
\chi_a(-\|w_i\|_{[0,t)}),
\chi_a(-\|w_j\|_{[0,t)}),c_a
\}\nonumber\\
&\qquad\ge
\min\{\chi_a(-\bar w),c_a\}
=d_a-d_{ma}.
\label{c_a_bound}
\end{align}
Since $p(0)\in\mathcal{D}_s$, substituting \eqref{c_a_bound} and the definition of $\mathcal{D}_s$ in \eqref{definition_D_s} into \eqref{h_ij_bound_example}, we have
\begin{align}
h_{ij}(p(t))\ge d_a-d_{ma}
\label{h_ij_safe}
\end{align}
for all $t\in[0,t_m)$. 

Combining \eqref{h_i_safe} and \eqref{h_ij_safe},
we obtain $p(t)\in\mathcal D_s$ for all
$t\in[0,t_m)$.

If $t_m\neq\infty$, then $p(t)$ tends to the boundary of $\mathcal{D}$ as $t\to t_m$ \cite[Theorem I.5.2]{hale1980ode}. For any $p(0)\in\mathcal{D}_s$, since $p(t)\in\mathcal{D}_s$ for all $t\in[0,t_m)$, $|p(t)|$ tends to $\infty$ as $t\to t_m$ whenever $t_m\neq \infty$. Under \eqref{apply_ui_definition_N} and \eqref{v_c_condition}, since $p(t)\in\mathcal{D}_s$ for all $t\in[0,t_m)$, there exists a positive constant $\bar{u}$ such that for each $i=1,\ldots,N$,
\begin{align}
|\dot{p}_i(t)|\le \bar{u}
\end{align}
for almost all $t\in[0,t_m)$. Therefore, if $t_m\neq\infty$, for each $i=1,\ldots,N$,
\begin{align}
|p_i(t)|\le |p_i(0)|+ \int_{0}^{t}|\dot{p}_i(\tau)|d\tau\le |p_i(0)|+ \bar{u}t_m,
\end{align}
which implies that $|p(t)|$ does not tend to $\infty$ as $t\to t_m$. This contradiction shows that $t_m=\infty$.

\section*{References}

\vspace{-5mm}

\bibliographystyle{IEEEtran}
\bibliography{reference}

\vspace{-10mm}

\begin{IEEEbiography}[{\includegraphics[width=1in, height=1.25in, clip, keepaspectratio]{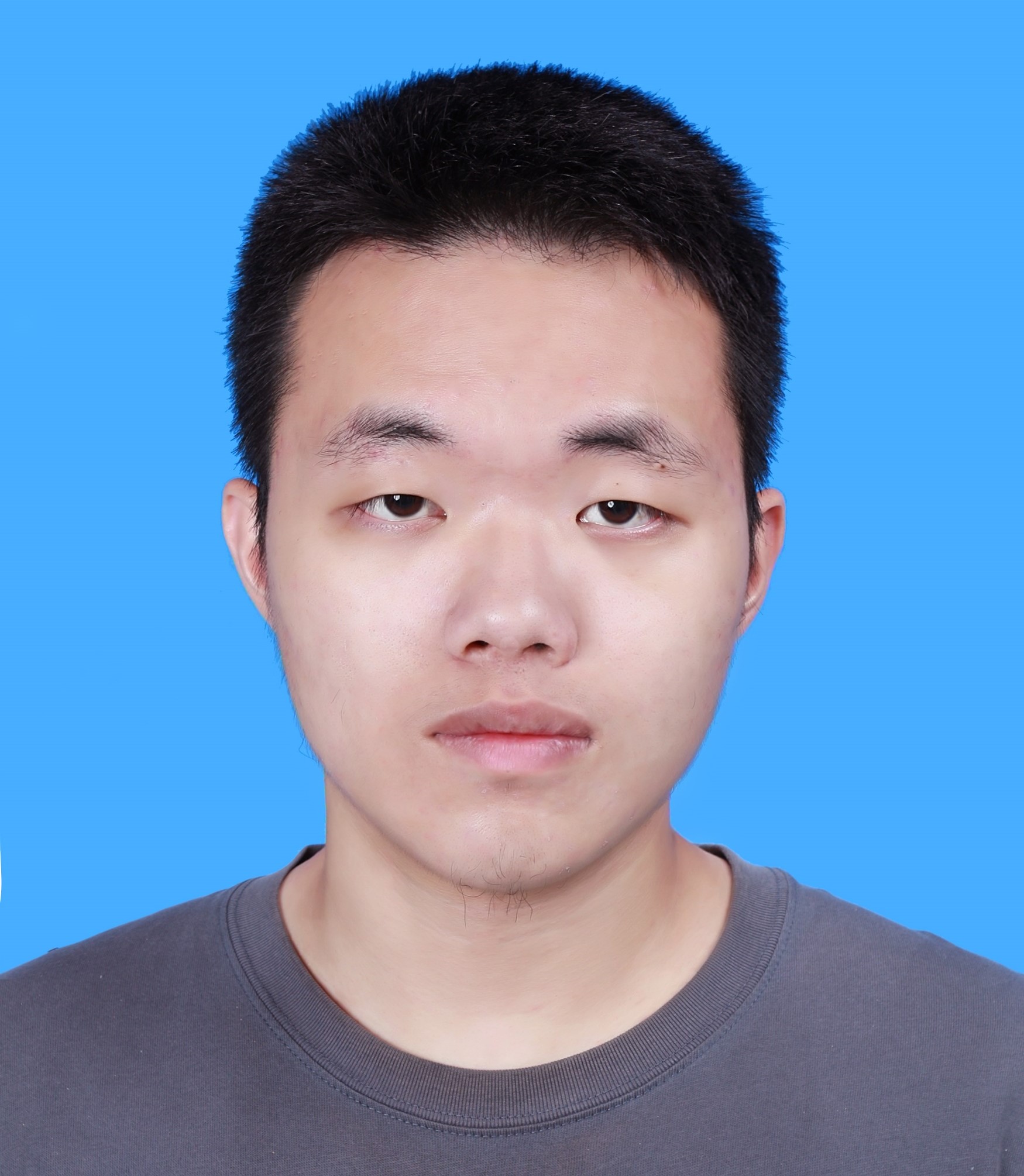}}]{Jianqiang Wang} received his B.E. degree in Automation from Northeastern University, China, in 2024. He is currently pursuing a Ph.D. in Control Science and Engineering at the same university. 
    
His research interests include stability theory and safety-critical control theory for interconnected nonlinear systems, and their applications to vehicles and robotic systems.
\end{IEEEbiography}

\vspace{-10mm}

\begin{IEEEbiography}[{\includegraphics[width=1in, height=1.25in, clip, keepaspectratio]{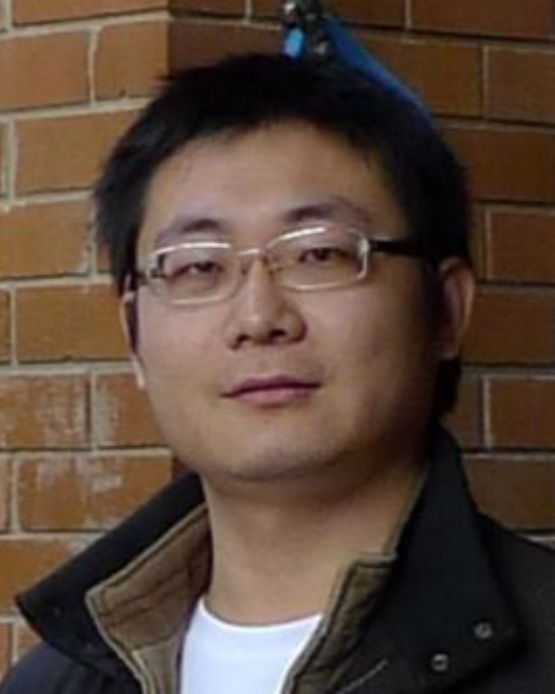}}]{Tengfei Liu} received his Ph.D. from the Australian National University, Canberra, Australia, in 2011. From 2011 to 2013, he was a postdoctoral fellow at the Polytechnic Institute of New York University. Since 2014, he has been with Northeastern University, China, where he is currently a full professor. 
    
His research interests focus on the stability and control of interconnected nonlinear systems. He has served as an Associate Editor for IEEE Transactions on Automatic Control since 2021.
\end{IEEEbiography}

\vspace{-10mm}

\begin{IEEEbiography}[{\includegraphics[width=1in,height=1.25in, clip,keepaspectratio]{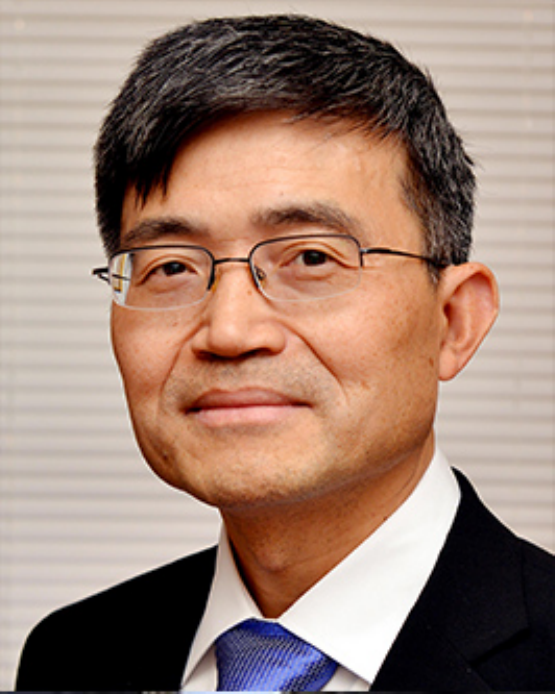}}]{Zhong-Ping Jiang} received the M.Sc. degree in statistics from the University of Paris XI, France, in 1989, and the Ph.D. degree in automatic control and mathematics from ParisTech-Mines, France, in 1993, under the direction of Prof. Laurent Praly.

Currently, he is an Institute Professor in the Department of Electrical and Computer Engineering at the Tandon School of Engineering, New York University. His main research interests include stability theory, robust/adaptive/distributed nonlinear control, robust adaptive dynamic programming, reinforcement learning and their applications to information, mechanical, transportation and biological systems. In these fields, he has written seven books and is the author/co-author of about 600 peer-reviewed journal and conference papers.

Prof. Jiang is currently serving on the IEEE Intelligent Transportation Systems Society’s Board of Governors, and has served as Deputy Editor-in-Chief, Senior Editor and Associate Editor for numerous journals. He is among the Clarivate Analytics Highly Cited Researchers and Stanford’s Top 2\% Most Highly Cited Scientists. In 2022, he received the Excellence in Research Award from the NYU Tandon School of Engineering. Prof. Jiang is a foreign member of the Academia Europaea (Academy of Europe) and an ordinary member of the European Academy of Sciences and Arts, and also is a Fellow of the IEEE, IFAC, CAA, AAIA and AAAS.

\end{IEEEbiography}

\end{document}